\documentclass[aps,amsmath,amssymb,floatfix,showpacs]{revtex4}
\usepackage{graphicx,color, bm} 
\usepackage{dcolumn} 

\usepackage{ulem} 

\newcommand{\be}{\begin{equation}}
\newcommand{\ee}{\end{equation}}

\begin{document}

\title{Long-time properties of MHD turbulence and the role of symmetries}

\author{Joshua E. Stawarz$^{1,2}$, Annick Pouquet$^1$ and Marc-Etienne Brachet$^3$}
\affiliation{$^1$\ Geophysical Turbulence Program, Computational and Information Systems Laboratory, National Center for Atmospheric Research, P.O. Box 3000, Boulder CO 80307, USA.\\
$^2$\ Laboratory for Atmospheric and Space Physics, University of Colorado at Boulder, Boulder, Colorado, USA \\
$^3$ \ Laboratoire de Physique Statistique de l'\'Ecole Normale
Sup{\'e}rieure, \\
associ{\'e} au CNRS et aux Universit{\'e}s Paris VI et VII,
24 Rue Lhomond, 75231 Paris, France.
}

\begin{abstract}
We investigate long-time properties of three-dimensional MHD turbulence in the absence of forcing and examine in particular the role played by the quadratic invariants of the system and by the symmetries of the initial configurations. We observe that, when sufficient accuracy is used, initial conditions with a high degree of symmetries, as in the absence of helicity, do not travel through parameter space over time whereas by perturbing these solutions either explicitly or implicitly using for example single precision for long times, the flows depart from their original behavior and can become either strongly helical, or have a strong alignment between the velocity and the magnetic field.  When the symmetries are broken,  the flows evolve towards different end states, as predicted by statistical arguments for non-dissipative systems with the addition of an energy minimization principle, as already analyzed in \cite{stribling_90} for random initial conditions using a moderate number of Fourier modes. Furthermore, the alignment properties of these flows, between velocity, vorticity, magnetic potential, induction and current, correspond to the dominance of two main regimes, one helically dominated and one in quasi-equipartition of kinetic and magnetic energy.
We also contrast the scaling of the ratio of magnetic energy to kinetic energy as a function of wavenumber to the ratio of eddy turn-over time to Alfv\'en time as a function of wavenumber. We find that the former ratio is constant with an approximate equipartition for scales smaller than the largest scale of the flow whereas the ratio of time scales increases with increasing wavenumber. 
\end{abstract}
\maketitle

\section{Introduction}\label{s:intro}
\subsection{The context}\label{ss:con}

Magnetic fields pervade the universe  and often play an essential role in configuring and constraining structures, as in the case of intergalactic jets, or closer to Earth, in the Solar Wind or the magnetosphere. 
Magnetic pressure contributes to the containment of the heliosphere \cite{opher_09}, and it may retard, together with turbulent pressure, the gravitational collapse of molecular clouds in the interstellar medium.
Magnetic fields are also known to accelerate the motion of charged particles  in the magnetospheres of the planets in the Solar System, as for example in the case of Jupiter's aurora \cite{saur} (see \cite{ergun1} for the main characteristic of the auroral emissions as of today).

Such magnetic fields have been observed in a variety of media to be turbulent, such as in the Solar Wind  \cite{goldstein} (see \cite{bruno} for a recent review), in the interstellar medium where it is thought to be responsible for strong velocity shear and intermittency \cite{falgarone},  or more recently in the heliosheath \cite{opher_11}. Using CLUSTER data with short separation between the satellites, it was shown in  \cite{sara} that  anisotropy of the energy Fourier spectra  develops at small scales as predicted in weak MHD turbulence theory \cite{galtier_00, galtier_02, galtier_rev}. Solar Wind turbulence can also help focus Langmuir wave packets which are routinely observed using Ulysses or STEREO spacecrafts \cite{hess}.

Furthermore, magnetic fields can lead to extreme energetic events due to reconnection of magnetic field lines in highly turbulent media; solar flares are one such example, the prediction of which is one of the purpose of space weather research because of the disturbance to Earth's communication networks and power grids. The penetration of the Solar Wind into the Earth's magnetosphere can be explained by the development of Kelvin-Helmoltz (KH) vortices, as observed \cite{hasegawa}, and a relationship between such KH instabilities and flux transfer events was found recently during substorms using multiple spacecrafts \cite{ericsson}.

An understanding of both fluid and MHD turbulence has escaped us for a long time. Is MHD turbulence similar to hydrodynamic turbulence, with a Kolmogorov energy spectrum, $E_{K41}(k)\sim \epsilon^{2/3} k^{-5/3}$ (hereafter K41), with $\epsilon\equiv dE/DT$ the energy dissipation rate, perhaps with an anisotropy due to the presence of strong uniform fields of magnitude $B_0$? Or is it different, because Alfv\'en waves propagate that alter and dampen the nonlinear dynamics of turbulent flows, leading to a so-called Irsohnnikov-Kraichnan energy spectrum, $E_{IK}(k)\sim [\epsilon B_0]^{1/2}k^{-3/2}$ (hereafter IK)?
And is there one answer to these questions, or is universality broken in MHD, as sometimes advocated? 
For example, it was found in \cite{lee} that one can observe three different energy spectra (K41, IK and weak turbulence, hereafter WT, 
$E_{WT}(k_{\perp})\sim k_{\perp}^{-2}$), for three different initial conditions of the magnetic field, using the same velocity field, and with the same ideal 
invariants, namely total energy, total magnetic helicity and total cross correlation between the velocity and the magnetic field, with 
moreover no imposed external field, no forcing, unit magnetic Prandtl number, and equal kinetic and magnetic energy initially? In 
other words, for such a set of initial conditions, nothing allows for distinguishing these three configurations from a statistical point of view, no externally imposed 
time-scale is present, and the only constraint is that numerically the flows and fields are forced to follow the four-fold symmetry of 
the initial conditions. Using this symmetry, one can gain in resolution and cost of computation, and thus the Reynolds numbers are quite large (Taylor Reynolds number in excess of 1200), with equivalent resolutions of $2048^3$ grid points. Similar results are found to hold in the forced case as well, for which long-time averaging is feasible \cite{giorgio}. 

The difference between these three power laws could be due to non-local interactions in Fourier space, between widely separated scales. Non-local interactions  are thought to be more prevalent in magnetohydrodynamics (MHD) than in hydrodynamics, as measured in high-resolution numerical simulations \cite{shell_I, shell_II}, and such non-locality in Fourier space is advocated in the differentiation between a K41 and an IK spectrum, but what would make one of the three flows studied in \cite{lee} more non-local than others? Perhaps the different behaviors come from another factor. On the one hand, it could be that the ratio of kinetic to magnetic energy, in particular in the gravest mode, matters, as indicated in \cite{lee}.
On the other hand, the invariants, which are quadratic in the basic fields, are identical but higher order moments could differ; for example, it was shown in \cite{gafd} that the skewness of one of the flows studied in \cite{lee} is measurably larger than that of the two other flows, when looking at both the velocity and the magnetic field.

The assumption that with the same invariants, the three flows should behave in similar ways is based on an assumption of ergodicity. However, the ergodicity of turbulent flows has been put into question in a variety of contexts. It has been observed that long-time memory effects can be found in such flows, for example in two-dimensional (2D) MHD turbulence \cite{gomez}, where large bursts of energy were observed to evolve on time scales of the order of one hundred turn-over times; it was also found more recently in numerical simulations of 3D hydrodynamic (HD) turbulence \cite{large_scale} and in laboratory experiments (see e.g. \cite{mouri} and references therein). In the atmospheric boundary layer, one observes that statistics can depend on the large scales; this  may be related to averaging over regions with local fluctuations in Reynolds number \cite{sreeni}.
Such a   transfer between non-local triads is shown to lead, however, to local exchanges of energy in hydrodynamics \cite{doma}.

The persistence of modes for long times in MHD, associated with large-scale coherent structures, was recently linked to normal modes appearing because of magnetic helicity, a large-scale invariant in the ideal case,
and leading to an apparent breaking of ergodicity at the largest scale, insofar as these structures persist for long times
\cite{shebalin_09}. A similar phenomenon occurs for hydrodynamics in the presence of solid body rotation: although the Coriolis force is linear, it affects the  dynamics of rotating turbulence in slowing it down substantially \cite{mininni}; this can be attributed to inertial waves, nonlinear transfer occurring only through (quasi-) resonances \cite{galtier_rev}.

In selective decay, some invariants are viewed as more sturdy than others; thus, they may influence the long-term dynamics of decaying turbulent flows. This hypothesis is based on the fact that invariants may have different physical dimensions, for example magnetic potential $<A^2>$, with ${\bf b}=\nabla \times {\bf A}$, and total energy $\frac{1}{2} <|{\bf v}|^2+|{\bf b}|^2>$ in 2D, or magnetic helicity $<{\bf a} \cdot {\bf b}>$ and total energy in 3D: since dissipation involves a Laplacian, it is thought that $<A^2>$ or $<H_M>$ will decay more slowly than energy. However, the third  invariant in ideal MHD, $H_C=<{\bf v} \cdot {\bf b}>$, has the same dimension as energy and it could also influence the long term dynamics, becoming strong in relative terms, that is with respect to the energy, implying an alignment between the velocity and the magnetic field, a phenomenon called  dynamic alignment. The relative importance of these two effects was explored, both theoretically and numerically in \cite{stribling_90, stribling_91}.  These theoretical considerations based on statistical mechanics of a truncated system of modes were backed up by rather low resolution numerical simulations which nevertheless clearly demonstrated the validity of the approach: the end state of such flows was determined by the respective ratio of their three invariants. Will the same happen here, when starting with the three flows studied in \cite{lee}, which statistically are equivalent but which display different inertial range dynamics at peak of dissipation (and in the statistically steady state as well)? This is the main question that this paper is addressing, using direct numerical simulations of the MHD equations in three space dimensions.

\subsection{The equations}\label{ss:eqs}

We now give the MHD equations for an incompressible fluid with $\bf {v}$ and $\bf {b}$ respectively the velocity and magnetic fields in Alfv\'enic units:
\begin{eqnarray}
&& \frac{\partial {\bf v}}{\partial t} + {\bf v} \cdot \nabla {\bf v} = 
    -\frac{1}{\rho_0} \nabla {\cal P} + {\bf j} \times {\bf b} + \nu \nabla^2 
    {\bf v} , 
\label{eq:MHDv} \\
&& \frac{\partial {\bf b}}{\partial t} = \nabla \times ( {\bf v} \times
    {\bf b}) +\eta \nabla^2 {\bf b} ;
\label{eq:MHDb}
\end{eqnarray}
$\rho_0=1$ is the (uniform) density (and ${\bf b}$ is then dimensionally a velocity as well, the Alfv\'en velocity), ${\cal P}$ is the total pressure,
${\bf \nabla} \cdot {\bf v} = \nabla \cdot {\bf b} = 0$, and $\nu$ and $\eta$ 
are respectively the kinematic viscosity and magnetic diffusivity;  we take $\nu=\eta$.
With 
$\nu=0,\ \eta=0$, the energy $E_T$, the cross helicity $H_C$ and the magnetic helicity $H_M$, defined as
$$E_T=E_V+E_M=\left<v^2+b^2\right>/2 \ , \ \ 
H_C=\left<{\bf v} \cdot {\bf b}\right>/2 \ , \ \ H_M=\left<{\bf A} \cdot {\bf b}\right>/2 \ , 
$$
are conserved.
Relative helicities can be defined as follows:
\begin{equation}
\rho_C= \cos{[\bf v}, {\bf b}] \ \ , \ \ \rho_M = \cos{[\bf A}, {\bf b}] \ \ , \ \ \rho_V= \cos{[\bf v}, {\mathbf \omega}] \ \ ;
\label{eq:relH} \end{equation}
they correspond to the degree of alignment between various vectors: the velocity, the magnetic field, the magnetic potential (with $\rho_M=\pm 1$ defining a force-free field), or the vorticity (with $\rho_V=\pm 1$ defining the so-called Beltrami configuration). 
In the latter case, the relative kinetic helicity involves the vorticity, ${\mathbf \omega}=\nabla \times {\bf v}$; the total kinetic helicity is an invariant of the Euler equations (${\bf b}\equiv 0, \ \nu\equiv 0$).
\vskip0.035truein

The kinetic energy spectrum is  the Fourier transform of the velocity two-point correlation function. Once homogeneity, isotropy and incompressibility have been taken into account, only two defining functions remain: $E_V(k)$ is proportional to the kinetic energy, with $\int E_V(k)dk =E_V=\frac{1}{2}\langle v^2 \rangle$, and the 
kinetic helicity, $H_V(k)$, stems from the anti-symmetric part of the velocity gradient tensor. Similar definitions hold for the magnetic and cross correlation functions (note that helicity is a pseudo-scalar). Finally, the kinetic and magnetic Reynolds numbers are defined as 
$$R_V=U_0L_0/\nu \ \ , \ \ R_M=U_0L_0/\eta \ ,$$
where $U_0,\ L_0$ are the characteristic velocity and length scale. The integral scale is defined as
$$
L_{int}=\frac{\int [E_V(k)/k]\ dk}{\int E_V(k)\ dk }\ .
$$


\subsection{Predictions from statistical mechanics} \label{ss:stat}

The statistical equilibria in 3D MHD were derived in \cite{frisch_75}. They are the long-time solutions to a truncated system of Fourier modes, with $k_{min}$ and $k_{max}$ the minimum and maximum wave numbers respectively; these modes are coupled through the nonlinear ideal MHD equations ($\nu \equiv 0, \ \eta \equiv 0$), and subject to the conservation of all quadratic invariants.
Defining $\alpha\not= 0,\ \beta$ and $\gamma$ as the Lagrange multipliers associated with the $E_T,\ H_M$ and $H_C$ invariants, namely
$\alpha E_T + \beta H_M + \gamma H_C      \ ,$
these equilibria read, assuming that the magnetic helicity is non-zero ($\beta\not= 0$):
 \begin{equation}
H_M(k) = - \frac{8\pi \beta} {\alpha^2 \Gamma^4}\  {\frac{1}{{\cal D}(k)}}  \ \ ; \ \   H_J(k)=k^2H_M(k) \ \ ; \ \  
H_C(k) =  \frac{\gamma \Gamma^2} {2\beta}\  {H_J(k)} \ \ , \ \  
H_V(k) =  \frac{\gamma^2} {4\alpha^2}\  {H_J(k)} \ , 
\label{eq:HJ} \end{equation}

\begin{equation}
\hskip-0.07truein E_M(k) =  -\frac{\alpha \Gamma^2 } {\beta }\  {H_J(k)}\ = \ \frac{8\pi k^2}{\alpha \Gamma^2} {\frac{1} {{\cal D}(k)}}\ \ , \ \ 
                E_V(k) = \left(\Gamma^2 {\cal D}(k) + \frac{\gamma^2}{4 \alpha^2} \right) E_M(k) {\  = \left(1- \frac{\beta^2}{4 \alpha^2 \Gamma^2}{ \frac{1} {k^2}} \right) E_M(k) \ ,}
\end{equation}
where $H_J=\int k^2 H_M(k)dk$ is the current helicity, and 
\begin{equation}
\alpha >0 \ ,  \hskip0.2truein \ \  \Gamma^2 = 1- \frac{\gamma^2}{4\alpha^2} > 0 \ ; \ \   \hskip0.2truein 
  {\cal D}(k) = \left(1-\frac{\beta^2}{\alpha^2 \Gamma^4} {\frac{1}{k^2}} \right) > 0 \   \ , \forall k \in [k_{min}, k_{max}] \ .
 \end{equation}
 $H_M$ and $H_C$ are not definite positive, and furthermore, $H_M$  does not have the same physical dimension as $E_T$ and $H_C$, and hence $\beta$ does not have the same physical dimension as $\alpha$ and $\gamma$. In order to fulfill realizability conditions (positivity of energy, and Schwarz inequalities involving the helicities), necessary relationships between coefficients  can be derived, involving $k_{min}$ (see \cite{frisch_75}).

When $\beta\equiv 0$ and thus $H_M(k)\equiv 0,\ H_J(k)\equiv 0$, one finds that the kinetic helicity is also equal to zero, that we have equipartition of energy at all wave numbers with $E_M(k)=E_V(k)=8\pi k^2/(\alpha \Gamma^2)$, and that $H_C(k)=-4\pi \gamma k^2/(\alpha^2 \Gamma^2)$; thus the relative cross helicity $2H_C(k)/E_V(k)$ is constant in that case. When $\gamma\equiv 0$ and thus $H_C(k)\equiv 0$, the kinetic helicity is also equal to zero and the kinetic energy has its non-helical expression, $E_V(k)=8\pi k^2/\alpha$;  the magnetic energy and helicity can peak at low wavenumber when $\beta$ is large enough, and the relative magnetic helicity $kH_M(k)/E_M(k)\sim 1/k$, i.e. it is stronger in the largest scales of the flow, a result that persists in the general case ($\beta\not=0, \gamma\not= 0$).

When considering $H_J(k)$ (instead of $H_M(k)$), note that all Fourier spectra are strictly proportional, with coefficients uniquely determined by initial conditions given the values of the invariants,  except for the kinetic energy; also note that, $\forall k$, one has $E_V(k)\le E_M(k)$, the equality arising only when there is either no magnetic helicity, or maximal cross-correlation, or for $k_{max}\rightarrow \infty$. Similarly, the residual helicity defined as
$$
H_R(k)=H_V(k)-H_J(k) = -\Gamma^2 H_J(k) \ , 
$$
 is of the sign opposite to that of the current and of the magnetic helicity, and $H_R(k)$ becomes equal to zero only for maximal cross-correlation ($\Gamma^2=0$), except for the trivial non-helical case of course. The relative helicity $H_R(k)$, integrated over the small scales, is the motor of the nonlinear dynamo problem, i.e. the growth of large-scale magnetic energy because of small-scale helical motions; note that $H_R$ reduces to the kinetic helicity in the kinematic regime when the magnetic field is weak, thus recovering the so-called ``alpha'' effect (see \cite{axel} for a recent comprehensive review).
 
It was shown in \cite{stribling_90, stribling_91} that these solutions can be seen as indicators of the long-time behavior of 3D MHD systems left to decay, because of a  principle of minimization of total energy. Three main regions of parameter space can be seen as attractors to the dynamics: a magnetic helicity dominated region, an alignment (strong $H_C$) region, and an intermediate region. 
The relaxation principle is well founded when there is magnetic helicity in the system, since dimensionally $H_M$ weighs the large scales more so than the energy or the cross-correlation, but when $H_M\equiv 0$, it is not so clear what happens. It can be shown, using minimum energy principles following \cite{woltjer}, that the resulting fields are ${\bf u}=0$ and ${\bf j}\propto {\bf b}$ when there is no cross-correlation (the constraint is simply that $H_M$ remain constant), whereas in the general case, the solution is a bit more involved (see eq. (3.13) {\it sq.} in \cite{houches}; see also \cite{stribling_91}).
The main purpose of this work is to investigate this long-term dynamics  when considering the three initial conditions used in \cite{lee} which have the same quadratic invariants ($E_T=1/4$, $H_M\equiv 0$ and $H_C \le 4\%$ in relative terms) and thus presumably the same final asymptotic state and yet, at peak of dissipation, show clear differences in their inertial range scaling. We shall also investigate other deterministic flows with either cross-helicity or magnetic helicity to see whether they evolve as well towards these attractors.

\subsection{Description of the initial conditions for all the computations} \label{ss:runs}

\begin{figure*} \begin{center}
\includegraphics[width=7.5cm, height=7.27cm ]{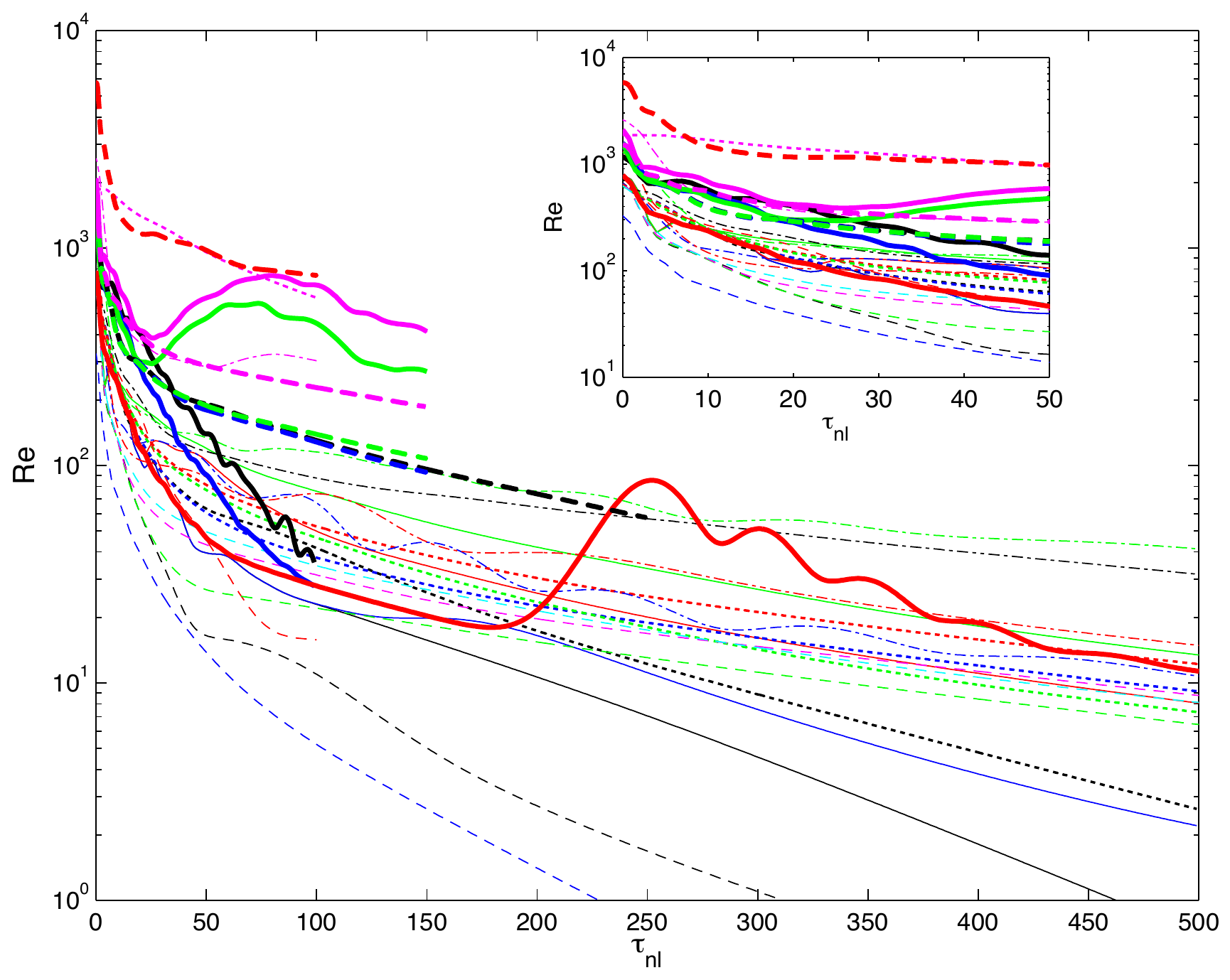}
\includegraphics[width=2.7cm, height=7.1cm]{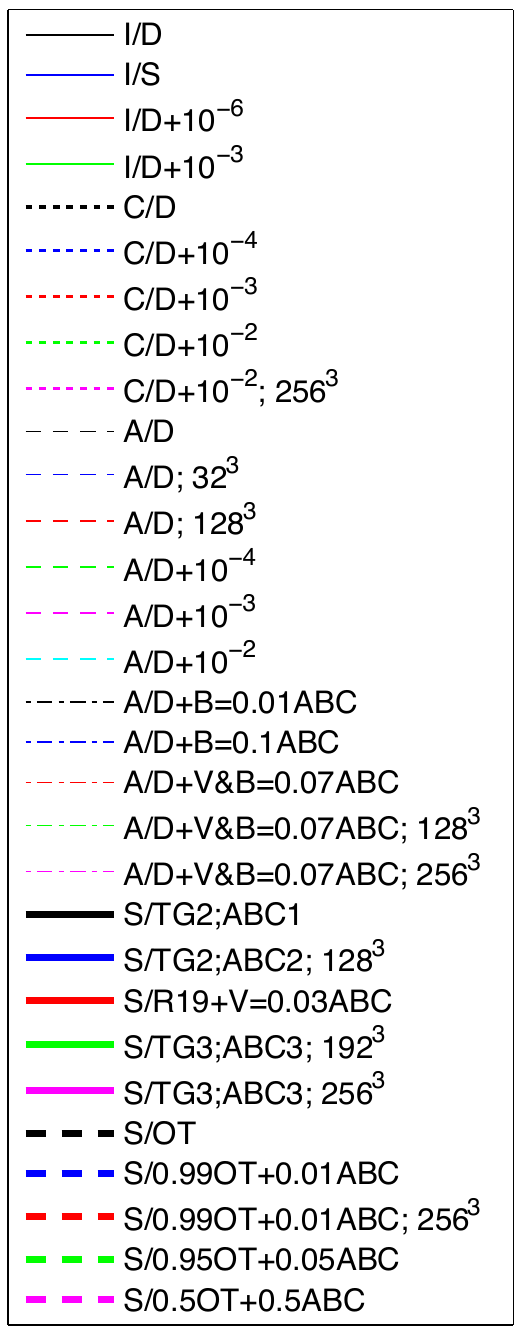}
\includegraphics[width=7.5cm, height=7.12cm ]{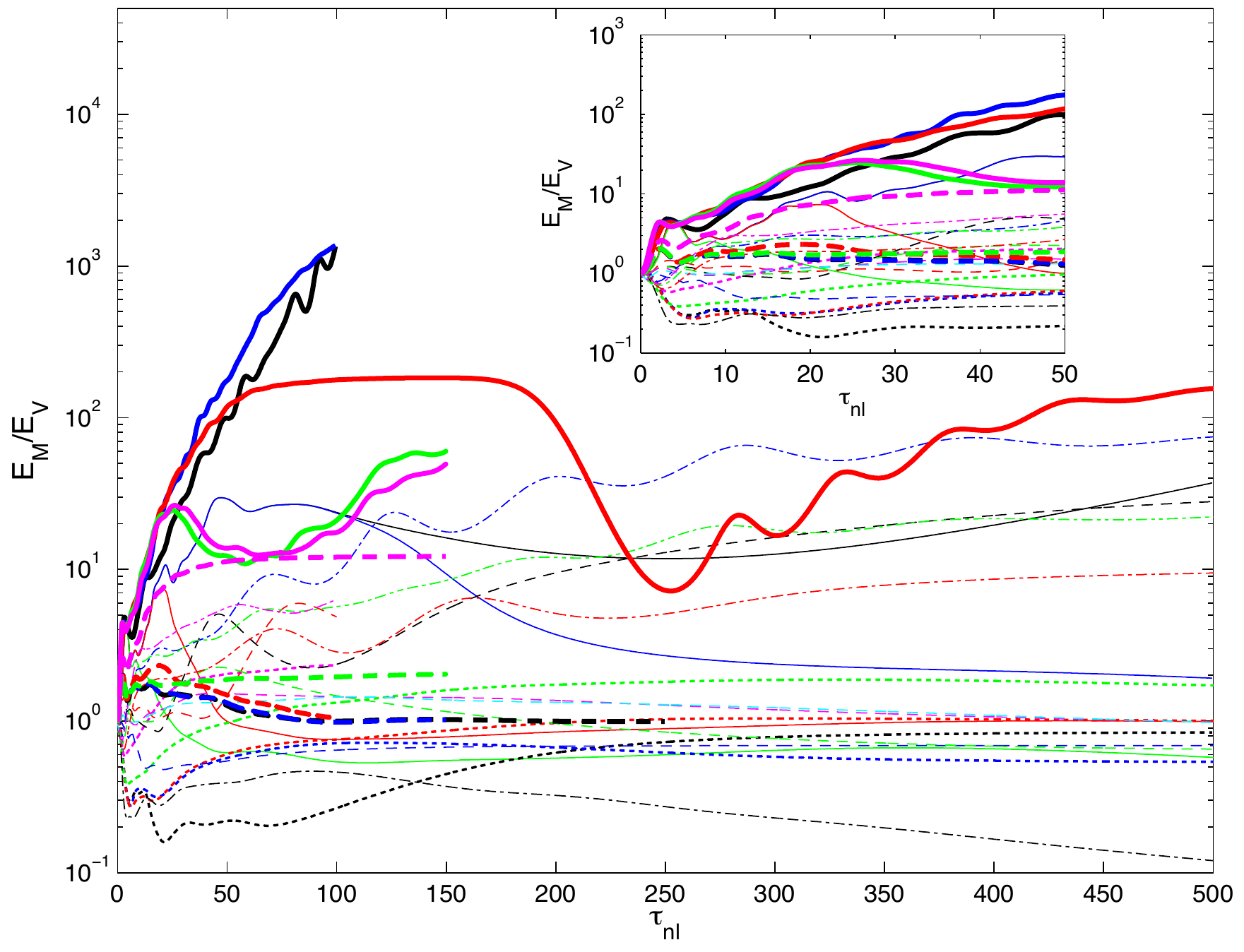}
 \caption{
 {\it Left:}  Evolution of the Reynolds number defined as 
 $Re \equiv v_{rms} L_{int}/ \nu$ where $L_{int}$ is the integral scale  
 (see text) as a function of time for most of the flows in Table \ref{tab1}. The plot is in lin-log coordinates and time is measured in units of turn-over time for each configuration. The inset shows a blow-up of the results for early times. Each run is indicated by symbols (and colors online) that are indicated in the center, a choice that is also followed in Figs. \ref{fig:map} and \ref{fig:rho} .
 {\it Right:} Magnetic to kinetic energy ratio for most of the runs in Table \ref{tab1} in lin-log coordinates. Many runs gather around quasi-equipartition, either above or below but for some runs, mostly those in which the relative amount of magnetic helicity grows significantly (such as runs R16 -- R21b), this ratio gets substantially larger than unity.
 }  \label{fig:ratio} \end{center}  \end{figure*}

\begin{figure*}  \begin{center}
\includegraphics[width=18cm]{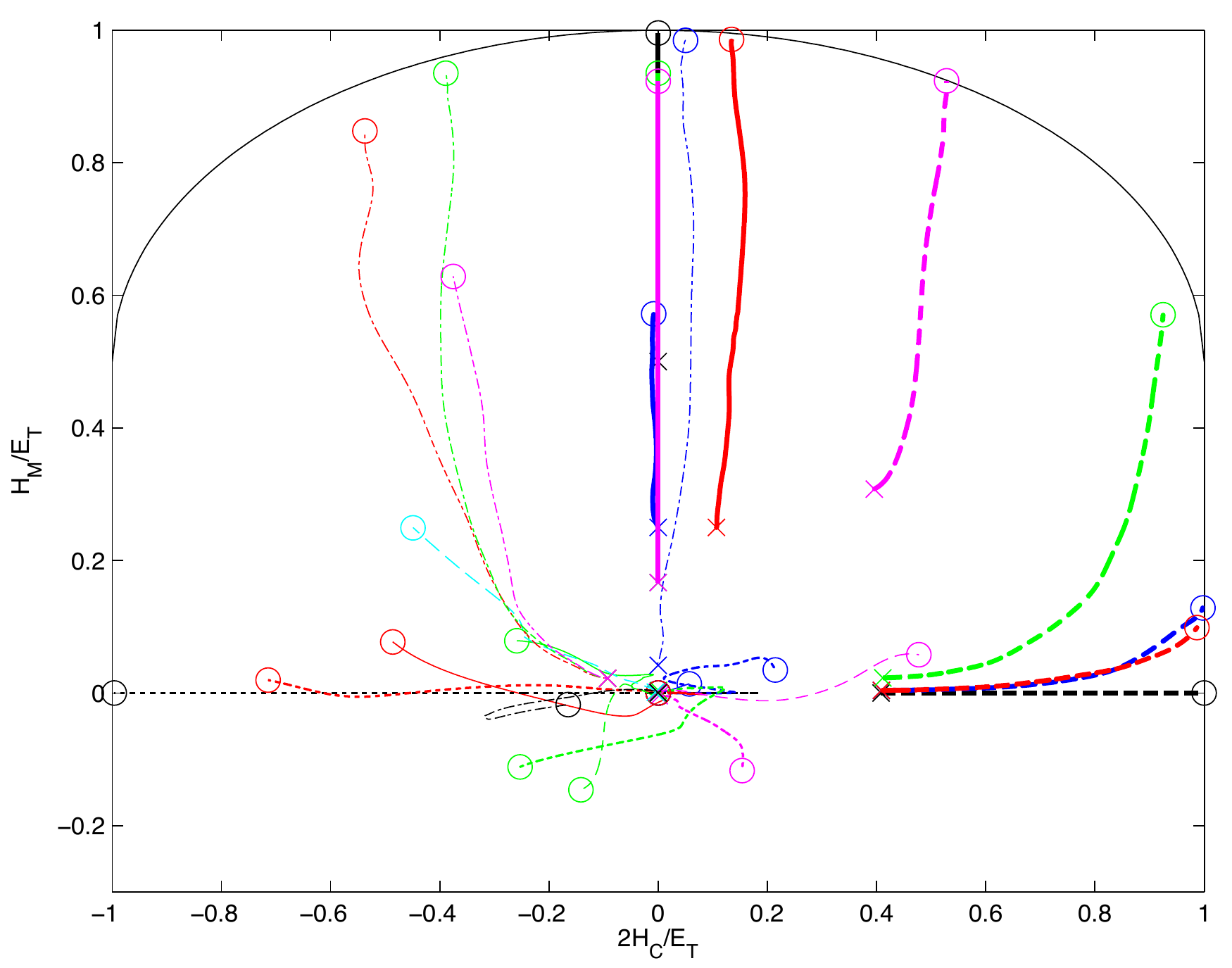}
 \caption{Map of the trajectories taken through the ($2H_C/E_T$, $H_M/E_T$) parameter space by the various runs in Table \ref{tab1}. 
The color scheme is the same as that given in Fig. \ref{fig:ratio} with ``x'' marking the start position of each run and ``o'' marking the position 
at the final time of the run. Three runs (R5a, R5b, and R6) follow the black dotted line which goes from the origin to the point (-1,0) and 
8 runs (R1a, R1b, R10a, R10b, R10c, R10d, R11a, and R11b) remain at the origin of the parameter space. Note that the origin of the parameter space is 
unstable and when symmetries are broken by perturbations the resulting trajectories can go in any direction depending of the phases of the perturbation.
The energy minimization principle predicts that flows will move towards the boundaries of this space; that is $2H_C/E_T = \pm 1$ when the 
end value of $H_M/E_T < 0.5$, and on the marked ellipsoidal curve bounding the upper portion of the plot when the end value of $H_M/E_T \ge 0.5$. 
The parameter space is symmetric for negative values of $H_M/E_T$. 
 }  \label{fig:map}  \end{center}   
 \end{figure*}

Table \ref{tab1} summarizes the main characteristics of the 36 runs described in this paper. Further details on the computations are given below, when specifying the initial velocity and magnetic fields.  All runs use the Geophysical High-Order Suite for Turbulence code (GHOST)  \cite{hybrid} unless otherwise stated in the ``Remarks'' column; TYGRS stands for a code which implements the four-fold symmetries of the Taylor-Green (TG) flow and its extensions to MHD \cite{lee}.
Values at $t=T_{100}=100 \tau_{NL}$ of 2$H_C/E_T$, $H_M/E_T$, and $E_T$, with $\tau_{NL}=L_0/U_{0}$, are given in the Table; note that $k_{min}=1$  has been taken as a normalizing factor for the ratio involving  magnetic helicity. 
 For all runs $\nu=\eta$ and $E_M=E_V=0.125$ initially. The groups divided by horizontal lines correspond to different line styles in  Figs. \ref{fig:ratio}, \ref{fig:map}, and \ref{fig:rho}.  All runs are performed on  a grid of $64^3$ points with $\nu=2\times 10^{-3}$, except for runs  
R10c and R11b  done on grids of $32^3$ points, runs R10d, R17b, and R19 done  on grids  
of $128^3$ points, run R21 done on a grid of $192^3$ points, and runs R9b and R23b done on grids 
of $256^3$ points (see the last column for the value of the viscosity in these cases). 
/D and /S stand for double and single precision. I, A and C in the second column refer to the Taylor-Green  
flows studied in \cite{lee}:  I, A and C are for the insulating boundary conditions (I), the alternate insulating conditions (A) and the conducting one (C); by insulating or conducting it is meant that in the box in which the computations are performed, the current is either parallel or normal to the walls. In most cases (except those labeled ``TYGRS''), the symmetries of the Taylor-Green initial conditions are not enforced and can be broken. 

Runs in which noise of amplitude $10^{-x}$ relative to the host flow has been added to both the kinetic 
and magnetic energy are denoted ``$+10^{-x}$''  in the second column. Furthermore,
``+V=xABC''/ or ``+B=xABC''/ or ``V\&B=xABC''  indicate that a Beltrami ABC flow has been added to either the velocity, magnetic field, 
or both for the initial conditions, with ``x'' indicating the fraction (in terms of energy) of the initial condition which is ABC. 
``TGx;ABCy'' stands for a modified Taylor-Green velocity at $k=x$ and a magnetic field which  
is a Beltrami ABC flow at $k=y$.
OT stands for the Orszag-Tang vortex generalized to three dimensions as the initial condition studied in \cite{PPS95}, and  
finally the last 4 runs have the velocity of \cite{PPS95} and a  
mixture of OT and ABC with the specified fractions for the magnetic field. 
The purpose here is to be able to vary the cross-helicity $H_C$ and the magnetic helicity $H_M$ of well-studied configurations in MHD turbulence, at a fixed total energy $E_T$, the same in all runs; indeed, all
  computations have equal initial kinetic and magnetic energy, with $E_T=E_V+E_M=1/4$.

GHOST is a general purpose pseudo-spectral community code with periodic boundary conditions;  the code is now parallelized up to $\sim 98,000$ processors, using a hybrid (MPI-Open-MP) methodology that becomes advantageous at high resolution \cite{hybrid}. Runs R1b, R5b, and R10b are done using a similar code, TYGRS, but in which the four-fold symmetries of the Taylor-Green configuration are enforced at all times \cite{brachet_TG, brachet_PRL}; TYGRS follows the same parallelization methodology as GHOST. The Taylor-Green velocity is:
$$
v_x^{TG} = v_0^{TG} \sin k_vx \ \cos k_vy \ \cos k_vz, \ \  v_y^{TG} = -v_0^{TG} \cos k_vx \ \sin k_vy \ \cos k_vz, \ \  v_z^{TG} = 0 \ ,
$$
and the three different initial conditions for the magnetic field are in that case:
$$
b_x^I = b_0^I \cos k_mx \ \sin k_my \ \sin k_mz, \ \  b_y^I = b_0^I \sin k_mx \ \cos k_my \ \sin k_mz, \ \  b_z^I = -2b_0^I \sin k_mx \ \sin k_my \ \cos k_mz \ ;
$$
$$
b_x^A= b_0^A \cos k_mx \ \sin k_my \ \sin k_mz, \ \  b_y^A = -b_0^A \sin k_mx \ \cos k_my \ \sin k_mz, \ \  b_z^A = 0 \ ; \ $$
and
$$
b_x^C = b_0^C \sin k_mx \ \cos k_my \ \cos k_mz, \ \  b_y^C = b_0^C \cos k_mx \ \sin k_my \ \cos k_mz,  \ \ b_z^C= -2b_0^C \cos k_mx \ \cos k_my \ \sin k_mz \ .
$$
When computations in which these initial fields  are perturbed with an added noise,  the amplitude of that noise relative to the energy in the Taylor-Green initial condition is indicated in the second column. This noise has randomly generated phases with an energy spectrum of the form:
$$
N=N_0 \exp \left( - \frac{(\log k - \log k_0)^2}{2 (\log \sigma)^2} \right) \ .
$$
In all cases the noise is centered around $k_0=2$ and has $\sigma=2$. Noise of this form is added to both the magnetic and kinetic energy and introduces small perturbations in the initial magnetic helicity and cross helicity relative to the total energy depending on the random phases generated and the amplitude of the noise.

We also performed some runs which have significant amounts of helicity since helicity is a main indicator of the behavior of such flows, at least in the ideal regime. Since the Taylor-Green runs have no helicity, different configurations are also studied. The ``TGx;ABCy'' type is one for which the velocity is a modified Taylor-Green, such that $v_x^{TG^\prime} = -v_y^{TG}$, $v_y^{TG^\prime}=-v_x^{TG}$, and $v_z^{TG^\prime}=0$, centered at wavenumber $k_v= x$ with $x$ equal to either 2 or 3, and the magnetic field is a Beltrami ABC flow centered at wavenumber $k_m=y$ with $y$ equal to either 1, 2, or 3; the ABC magnetic field is:

\begin{eqnarray}
b_x^{ABC} &=&  b_0^{ABC} \left[B \cos(k_m y) + C \sin(k_m z) \right] \nonumber  \ , \\
b_y^{ABC} &=& b_0^{ABC}  \left[A \sin(k_m x) + C \cos(k_m z) \right] \nonumber \ ,  \\ 
b_z^{ABC} &=& b_0^{ABC} \left[A \cos(k_m x) + B \sin(k_m y) \right] \nonumber \ .
\end{eqnarray}


The OT configuration is that of the generalization of the Orszag-Tang vortex to three dimensions, as studied in \cite{PPS95}, with the velocity and magnetic fields defined as:
$$
v_x^{OT} = -2v_0^{OT} \sin k_vy, \ \  v_y^{OT} = 2v_0^{OT}  \sin k_vx , \ \  v_z^{OT} = 0 \ ,
$$
and
$$
b_x^{OT} = 
b_0^{OT} \left[-2 a \sin 2k_my  + a \sin k_mz \right], \ \  
b_y^{OT} = 
b_0^{OT} \left[ 2 a \sin k_mx  + a \sin k_mz \right], \ \ 
b_z^{OT} = b_0^{OT} \left[ a \sin k_mx  + a \sin k_my \right] \ ;
$$
 the parameter $a$ allows one to modify the cross-correlation between the two fields; the choice $a=0.8$ gives a relative correlation of $0.41$. 

Finally, initial conditions which are mixtures of the above types are also studied. Runs R15 and R16 have the Taylor-Green velocity and perturb the Taylor-Green magnetic field initial condition with an ``A''  configuration, with an ABC Beltrami field such that:
\begin{eqnarray}
b_x^{A+ABC} &=& b_0^{A+ABC} \left( \xi_1 \cos k_{m1}x \ \sin k_{m1}y \ \sin k_{m1}z + \xi_2 \left[B \cos(k_{m2} y) +  C \sin(k_{m2} z) \right] \right)  \nonumber  \ , \\
b_y^{A+ABC} &=& b_0^{A+ABC} \left( -\xi_1 \sin k_{m1}x \ \cos k_{m1}y \ \sin k_{m1}z +  \xi_2 \left[A \sin(k_{m2} x) + C \cos(k_{m2} z) \right] \right) \nonumber  \ , \\
b_z^{A+ABC} &=& b_0^{A+ABC} \xi_2 \left[ A \cos(k_{m2} x) + B \sin(k_{m2} y) \right] \nonumber  \ .
\end{eqnarray}
The parameters $\xi_1$ and $\xi_2$ set the relative fractions of the Taylor-Green and ABC portions of the initial condition. Run R15 has $\xi_1=0.99$ and $\xi_2=0.01$ and run R16 has $\xi_1=0.9$ and $\xi_2=0.1$. Both of these flows have $k_{m1}$ and $k_{m2}$ such that both portions of the initial condition are at $k=3$. Initial conditions such as these allow for a perturbation in the magnetic helicity without significantly perturbing the cross helicity. 
Runs R17a,  R17b, and R17c use the above combined magnetic field, but also have a velocity that combines the Taylor-Green and ABC flows in a similar fashion:
\begin{eqnarray}
v_x^{TG+ABC} &=& v_0^{TG+ABC} \left( \xi_1 \sin k_{v1}x \ \cos k_{v1}y \ \cos k_{v1}z + \xi_2 \left[B \cos(k_{v2} y) +  C \sin(k_{v2} z) \right] \right)  \nonumber  \ , \\
v_y^{TG+ABC} &=& v_0^{TG+ABC} \left( -\xi_1 \cos k_{v1}x \ \sin k_{v1}y \ \cos k_{v1}z +  \xi_2 \left[A \sin(k_{v2} x) + C \cos(k_{v2} z) \right] \right) \nonumber \ ,  \\
v_z^{TG+ABC} &=& v_0^{TG+ABC} \xi_2 \left[ A \cos(k_{v2} x) + B \sin(k_{v2} y) \right] \nonumber \ .
\end{eqnarray}
In the case of the three R17 runs, both the magnetic field and velocity initial conditions have $\xi_1=0.93$ and $\xi_2=0.07$. The magnetic field is such that both portions of the initial condition are at $k=3$, but the velocity has the Taylor-Green portion of the flow at $k=2$ and the ABC portion at $k=3$ initially. This allows for a perturbation in both the magnetic helicity and cross helicity.
R20 involves a velocity which is a combination of the modified Taylor-Green velocity and a Beltrami ABC flow, such that: 
\begin{eqnarray}
v_x^{TG^{\prime}+ABC} &=& v_0^{TG^{\prime}+ABC} \left( \xi_1 \cos k_{v1}x \ \sin k_{v1}y \ \cos k_{v1}z + \xi_2 \left[B \cos(k_{v2} y) +  C \sin(k_{v2} z) \right] \right)  \nonumber  \ , \\
v_y^{TG^{\prime}+ABC} &=& v_0^{TG^{\prime}+ABC} \left( -\xi_1 \sin k_{v1}x \ \cos k_{v1}y \ \cos k_{v1}z +  \xi_2 \left[A \sin(k_{v2} x) + C \cos(k_{v2} z) \right] \right) \nonumber  \ , \\
v_z^{TG^{\prime}+ABC} &=& v_0^{TG^{\prime}+ABC} \xi_2 \left[ A \cos(k_{v2} x) + B \sin(k_{v2} y) \right] \nonumber \ .
\end{eqnarray}
In run R20, $\xi_1=0.97$, $\xi_2=0.03$, and $k_{v1}$ and $k_{v2}$ are set such that both the Taylor-Green and ABC porions of the flow are at $k=2$. When combined with an ABC magnetic field this results in the addition of cross-correlation between the velocity and magnetic fields of $0.11$ relative to the total energy.

The type ``$\xi_1$ OT+$\xi_2$ ABC'' is an initial condition for the velocity which is the OT vortex and combines the OT and ABC magnetic fields, such that:
\begin{eqnarray}
b_x^{OT+ABC} &=& b_0^{OT+ABC} \left( \xi_1 \left[ -2 a \sin 2k_{m1}y  + a \sin k_{m1}z \right] + \xi_2 \left[ B \cos(k_{m2} y) +  C \sin(k_{m2} z) \right ]\right)  \nonumber  \ , \\
b_y^{OT+ABC} &=& b_0^{OT+ABC} \left( \xi_1 \left[ 2 a \sin k_{m1}x  + a \sin k_{m1}z \right] +  \xi_2 \left[ A \sin(k_{m2} x) + C \cos(k_{m2} z) \right] \right) \nonumber \ ,  \\
b_z^{OT+ABC} &=& b_0^{OT+ABC} \left( \xi_1 \left[ a \sin k_{m1}x  + a \sin k_{m1}y \right] +  \xi_2 \left[ A \cos(k_{m2} x) + B \sin(k_{m2} y) \right] \right) \nonumber \ ,
\end{eqnarray}
where $\xi_1$ and $\xi_2$ set the relative fractions of OT and ABC respectively. Each of the flows of this type are such that $k_{m1}=k_{m2}=1$.

Some runs were performed for more than 1000 $\tau_{NL}$, where $\tau_{NL}=L_0/U_{0}$ is the turn-over time, and the maximum number of modes in the largest runs on grids of $256^3$ points is in excess of one million. 

 \begin{turnpage} \begin{table}
\caption{Initial conditions of the 36 runs,  data at $t=100 \tau_{NL}$ and at the final time $T_F$; three runs are done with TYGRS (see text).
The runs are performed on grids of $64^3$ points (but see ``Remarks''). 
Horizontal lines divide groups with different line styles in Figs. \ref{fig:ratio}, \ref{fig:map} \& \ref{fig:rho}. 
/D or /S stand for double or single precision.  I, A \& C are for the TG flows \cite{lee};  ``$+10^{-x}$'' indicates  runs with 
noise of amplitude $10^{-x}$ relative to the host flow; 
``+V=xABC''/``+B=xABC''/``V\&B=xABC''  stand for a Beltrami (ABC) flow added to either ${\bf v}$, ${\bf b}$
or both, ``x'' being the energy fraction  which is ABC. ``TGx;ABCy'' is a modified TG velocity at $k=x$ with ${\bf b}$ an ABC field at $k=y$. OT is the configuration studied in \cite{PPS95}, and  
 the last 4 runs have ${\bf v}$ as in \cite{PPS95}, and a  
mixture of OT and ABC with the specified fractions for ${\bf b}$. 
}
\noindent\makebox[\textwidth]{  \begin{tabular}{cccccccccccccccc}  
\hline \hline
                        & & \multicolumn{2}{c}{Initial Values}&                &   
&                \multicolumn{3}{c}{Values at $T_{100}=100\tau_{NL}$ } &&&  
\multicolumn{4}{c}{Values at $T_{F}/\tau_{NL}$}\\
                                \cline{3-4}                                                         \cline{7-10}                                     
                      \cline{12-15}
Run        & Type &   2$H_C/E_T$         &     $H_M/E_T$          &$k_{V}$ &  $k_{M}$&  
$E_M/E_V$&    2$H_C/E_T$ &    $H_M/E_T$         & $10^4 E_T$ &$T_F$ &  
$E_M/E_V$&    2$H_C/E_T$ &    $H_M/E_T$         & $10^6 E_T$          & Remarks        \\
\hline \hline
R1a      & I/D           & 0.00	   & 0.00    & 2       & 2       & 23.7      & 0.00        & 0.00      & 30.9      
        & 500           & 37.2    & 0.00    & 0.00    & 3.84    &            \\
R1b     & I/D/T         & 0.00     & 0.00    & 2       & 2       & 23.7      & 0.00        & 0.00      & 30.9  
        & 500           & 37.2    & 0.00    & 0.00    & 3.84    & TYGRS     \\
R2     & I/S           & 0.00     & 0.00    & 2       & 2       & 23.5      & 0.00        & 0.00      & 30.9
        & 500           & 1.91     & 0.06    & 0.01    & 2.26    &            \\
R3      & I/D+$10^{-6}$   & 0.00     & 0.00    & 2       & 2       & 0.76       & -0.01       & -0.02     & 6.18 
        & 500           & 0.99     & -0.48   & 0.08    & 7.49    &            \\
R4      & I/D+$10^{-3}$   & 0.00     & 0.00    & 2       & 2       & 0.54       & -0.02       & 0.02      & 7.83 
        & 500           & 0.58     & -0.26   & 0.08    & 16.2    &            \\
\hline
R5a      & C/D           & 0.00     & 0.00    & 2       & 3       & 0.27       & -0.75       & 0.00      & 4.80 
        & 500           & 0.84     & -1.00   & 0.00    & 2.56    &            \\
R5b     & C/D/T         & 0.00     & 0.00    & 2       & 3       & 0.27       & -0.75       & 0.00      & 4.80
        & 500           & 0.84     & -1.00   & 0.00    & 2.56    & TYGRS     \\
R6     & C/S           & 0.00     & 0.00    & 2       & 3       & 0.27       & -0.75       & 0.00      & 4.80  
        & 500           & 0.84     & -1.00   & 0.00    & 2.56    &            \\
R7      & C/D+$10^{-4}$   & 0.00     & 0.00    & 2       & 3       & 0.71        & 0.05       & 0.03      & 2.90
        & 500           & 0.54     & 0.21    & 0.03    & 6.89    &            \\
R8      & C/D+$10^{-3}$   & 0.00     & 0.00    & 2       & 3       & 0.75       & -0.22       & 0.01      & 4.64  
        & 500           & 1.00     & -0.71   & 0.02    & 16.0    &            \\ 
R9a    & C/D+$10^{-2}$   & 0.00     & 0.00    & 2       & 3       & 1.32       & 0.06        & -0.03     & 4.69
        & 500           & 1.71     & -0.25   & -0.11   & 8.27    &            \\
R9b    & C/D+$10^{-2}$ & 0.00     & 0.00    & 2       & 3       & 2.33       & 0.15       & -0.12     & 8.98
        & 100           & --     & --   & --   & --    &   $256^3$; $\nu=6.67\times 10^{-4}$         \\
\hline
R10a     & A/D           & 0.00     & 0.00    & 2       & 3       & 2.36       & 0.00        & 0.00      & 1.14  
        & 500           & 28.0    & 0.00    & 0.00    & 0.36    &            \\
R10b    & A/D/T         & 0.00     & 0.00    & 2       & 3       & 2.36       & 0.00        & 0.00      & 1.14
        & 500           & 28.0    & 0.00    & 0.00    & 0.36    & TYGRS     \\
R10c     & A/D   & 0.00     & 0.00    & 2       & 3       & 0.64       & 0.00        & 0.00      & 0.40
        & 1000          & 0.69     & 0.00    & 0.00    & 1.21$\times 10^{-8}$ & $32^3$; $\nu=4\times 10^{-3}$     \\
R10d     & A/D   & 0.00     & 0.00    & 2       & 3       & 4.86       & 0.00        & 0.00      & 1.50
        & 100           & --       & --      & --      & --      & $128^3$; $\nu=1\times 10^{-3}$    \\
R11a    & A/S           & 0.00     & 0.00    & 2       & 3       & 2.36       & 0.00        & 0.00      & 1.14 
        & 500           & 28.0    & 0.00    & 0.00    & 0.36    &            \\
R11b     & A/S   & 0.00     & 0.00    & 2       & 3       & 0.64       & 0.00        & 0.00      & 0.40
        & 1000          & 0.69     & 0.00    & 0.00    & 1.21$\times 10^{-8}$ & $32^3$; $\nu=4\times 10^{-3}$     \\
R12     & A/D+$10^{-4}$   & 0.00     & 0.00    & 2       & 3       & 1.47       & -0.07       & 0.01      & 1.50   
        & 500           & 0.65     & -0.14   & -0.15   & 3.73    &            \\
R13     & A/D+$10^{-3}$   & 0.00     & 0.00    & 2       & 3       & 1.45       & 0.22        & -0.01     & 2.35  
        & 500           & 0.98     & 0.48    & 0.06    & 8.08    &            \\
R14     & A/D+$10^{-2}$   & -0.01    & 0.00    & 2       & 3       & 1.44       & -0.24       & 0.08      & 2.84  
        & 500           & 0.98     & -0.45   & 0.25    & 7.36    &            \\
\hline
R15     & A/D+B=0.01ABC     & 0.00     & 0.00    & 2       & 3       & 0.46       & -0.26       & -0.02     & 7.99
        & 500           & 0.12     & -0.16   & -0.02   & 57.7    &            \\
R16	& A/D+B=0.1ABC      & 0.00     & 0.04    & 2       & 3       & 9.30       & 0.05        & 0.85      & 25.8
        & 500           & 74.81    & 0.05    & 0.98    & 443     &            \\
R17a     & A/D+V\&B=0.07ABC     & -0.09     & 0.02    & 2,3       & 3       & 2.86       & -0.52      & 0.55     & 12.3
        & 500           & 9.45     & -0.54   & 0.85   & 120    &            \\
R17b     & A/D+V\&B=0.07ABC     & -0.09    & 0.02    & 2,3       & 3       & 5.68       & -0.40       & 0.61     & 19.4
        & 500           & 22.2     & -0.39   & 0.94   & 511    &  $128^3$; $\nu=1\times 10^{-3}$          \\
R17c     & A/D+V\&B=0.07ABC     & -0.09    & 0.02    & 2,3       & 3       & 6.34      & -0.38       & 0.63     & 37.0
        & 100           & --    & --  & --   & --    &  $256^3$; $\nu=5\times 10^{-4}$          \\
\hline \hline
R18     & S/TG2;ABC1      & 0.00     & 0.50    & 2       & 1       & 1428       & 0.00        & 1.00      & 817 
        & 100           & --       & --      & --      & --      &            \\
R19     & S/TG2;ABC2      & 0.00     & 0.25    & 2       & 2       & 1379       & -0.01       & 0.57      & 535     
        & 100           & --       & --      & --      & --      & $128^3$; $\nu=1\times 10^{-3}$     \\
R20     & S/R19+V=0.03ABC & 0.11     & 0.25    & 2       & 2       & 177        & 0.15        & 0.57      & 282             
        & 500           & 155      & 0.13    & 0.99    & 1018    &            \\
R21a     & S/TG3;ABC3      & 0.00     & 0.17    & 3       & 3       & 20.5      & 0.00        & 0.85      & 316     
        & 150           & 60.2    & 0.00    & 0.94    & 2.71$\times 10^5$    & $192^3$; $\nu=6.67\times 10^{-4}$     \\
R21b     & S/TG3;ABC3      & 0.00     & 0.17    & 3       & 3       & 17.4      & 0.00        & 0.80      & 363     
        & 150          & 49.7    & 0.00    & 0.92   & 3.03$\times 10^5$    & $256^3$; $\nu=5\times 10^{-4}$    \\
\hline
R22     & S/OT             & 0.41     & 0.00    & 1       & 1,2                 & 1.00       & 0.99        & 0.00      & 21.0  
        & 250            & 0.99     & 1.00    & 0.00    & 337                 &            \\
R23a     & S/0.99OT+0.01ABC & 0.41     & 0.00    & 1       & 1,2                 & 0.98       & 0.99        & 0.12      & 19.7 
        & 150            & 1.02     & 1.00    & 0.13    & 957                 &            \\
R23b     & S/0.99OT+0.01ABC & 0.41     & 0.00    & 1       & 1,2                 & 1.05       & 0.99       & 0.10      & 62.4 
        & 100            & --     & --    & --    & --                 & $256^3$; $\nu=5\times 10^{-4}$            \\
R24     & S/0.95OT+0.05ABC & 0.41     & 0.02    & 1       & 1,2                 & 1.95       & 0.92        & 0.51      & 32.0 
        & 150            & 2.03     & 0.92    & 0.57    & 1863                &            \\
R25     & S/0.5OT+0.5ABC   & 0.40     & 0.31    & 1       & 1,2                 & 12.03      & 0.53        & 0.92      & 344 
        & 150            & 12.16    & 0.53    & 0.92    & 2.30$\times 10^4$   &            \\

\hline \hline  \label{tab1} \end{tabular}} \end{table} \end{turnpage} 

\subsection{Global properties for all the runs}

We show in Fig. \ref{fig:ratio} the temporal evolution of the Reynolds number (left) and of the ratio of magnetic to kinetic energy (right) for most of the  runs. The color table and symbols for runs is also given in Fig. \ref{fig:ratio}.
Since the runs are performed at relatively modest Reynolds numbers and numerical resolutions, but for long times, the Reynolds numbers eventually enter a regime of exponential decay where nonlinearities are weak. The burst of energy for run R20 at $t\sim 250$ is associated with the end of a plateau in the ratio $E_M/E_V$ and with a weak Lamb vector ${\cal L}={\bf v} \times {\mathbf \omega}$ (see Fig. \ref{fig:rho} below). Examining the energy ratio, it is clear that two main regimes develop in these runs: some are close to equipartition, with a tendency to have an excess in magnetic energy as predicted by the statistical ensembles, and one where the magnetic energy wins all, and presumably under the influence of a strong relative magnetic helicity and an accumulation of $H_M$ at the gravest mode of the computation.
 The run done on a grid of $32^3$ points has its Reynolds number getting too low and as a result behaves considerably differently from the other Taylor-Green ``A'' flow runs on grids of $64^3$ and $128^3$ points.
 {Of course ${\bf v}=0$ is a possible solution of the MHD equations; this corresponds to the hydrodynamic attractor which can also be fluid in the forced case when the magnetic Reynolds number is too low. }
 Also note that run R15, which is the Taylor-Green A flow perturbed by 1\% ABC magnetic field, moves towards a kinetically dominated state;
  by the end of the run, this flow has kinetic energy dominating over magnetic energy by approximately a factor of 10 in the gravest mode.

 \begin{figure*} \begin{center}
\includegraphics[width=10.0cm   ]{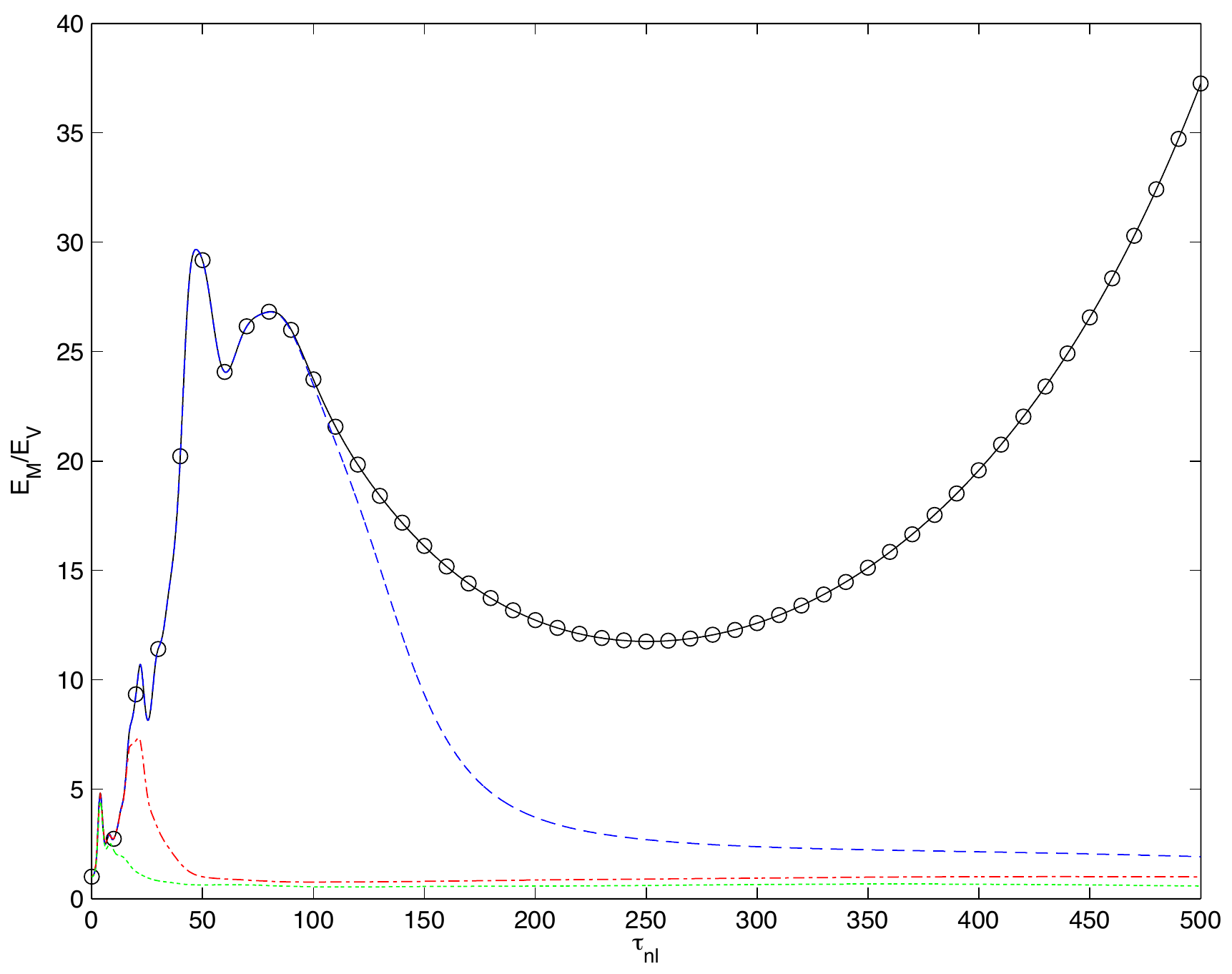}
 \caption{Comparison for the magnetic to kinetic energy ratio for the Taylor-Green I flow between two codes, one implementing symmetries (run R1b, black circles), the other one not, (run R1a, black solid line), both in double precision, 
 together with another run using single precision (run R2, blue dashed line).
 The double-precision in GHOST is sufficient to maintain the symmetries up to T=500, in units of turn-over times. Note the excellent agreement between the three computations performed on a grid of $64^3$ points up to $T\sim 100 \tau_{NL}$, time after which the single-precision run departs from the others and evolves towards quasi-equipartition (at the final time, $E_M/E_V\sim 2$, as often observed in the Solar Wind \cite{goldstein}).
The red dashed-dotted line and the green dotted line are the Taylor-Green I flow with added random noise of relative amplitude $10^{-6}$ (run R3) and $10^{-3}$ (run R4) respectively. Note that added noise behaves similarly to the single precision computation, but departs from the other runs and goes towards quasi-equipartition at earlier times.
  }  \label{fig:accu} \end{center}  \end{figure*}

\section{The role of accuracy and symmetries} \label{s:symm}

The ensemble of runs analyzed in this paper  is shown in Fig. \ref{fig:map} in a plane first introduced in \cite{stribling_90, stribling_91}; it delineates, in terms of the total energy, the relative importance of the two helical invariants. A peculiar feature of the I and A Taylor-Green runs is that, unless perturbed, they stay where they started, even though in these runs the symmetries are not imposed at all times. This may be related to the fact that it can be shown that, in the context of the fluid equations, symmetries are preserved by the dynamical evolution, a result that one may be able to extend to the MHD case \cite{bardos}. Unperturbed, these two flows do not evolve in parameter space at these low Reynolds numbers. In the presence of perturbations, they do cover parameter space and evolve towards configurations with either strong $H_M$ (and thus high ratio $E_M/E_V$, as in the case of runs R16, R17a, and R17b), or strong $H_C$ with near equipartition of kinetic and magnetic energy. Note that, in single precision and for long times, the I flow is perturbed by the accumulated round-off errors and it evolves toward another attractor, as shown in Fig. \ref{fig:accu}: whereas the accurate computation which maintains all symmetries evolves toward presumably
a magnetically-dominated Taylor state, the errors introduced by insufficient precision lead to a quasi-equipartition of energy. This same behavior, where the single precision computation evolves towards a different attractor after sufficiently long times, is not evident in either the A or C Taylor-Green flows; however similar effects are observed when random noise is explicitly added to these initial conditions.

The C flow also exhibits unique behavior as compared to the I and A flows in that, over the course of the computation, the accurate double precision run does not remain at the origin of the plane and instead moves to a state with $2H_C/E_T = -1$ and $H_M/E_T = 0$. As a result, the C flow without perturbations reaches an equipartitioned state, as opposed to a magnetically dominated state. 

We have also performed a more controlled and specific perturbation of the Taylor-Green symmetries by adding a fraction of a Beltrami ABC flow to the magnetic field and/or velocity of the Taylor-Green initial condition in the A configuration, i.e. by perturbing the flow explicitly with non-zero helicity (Runs 15--17c and 18--21b). By varying the amount of helical--ABC relative to non--helical Taylor-Green in the magnetic field, the value of $H_M/E_T$ can be adjusted in a controlled manner at $t=0$. By also adding a fraction of ABC to the velocity, a set amount of $2H_C/E_T$ can additionally be introduced to the flow.  With larger perturbations to the magnetic helicity and cross helicity, such as in runs R16 -- R17c, the symmetries are clearly broken and the runs reach the boundaries of the parameter space as predicted by the minimum energy principle (see Fig. \ref{fig:map}). Note that run R17c, which is performed on a grid of $256^3$ points, is only run for $100 \tau_{NL}$. If this run were continued to longer times, as the other two R17 runs are, this run would likely reach the boundary. In the case of run R16, only the magnetic field  initial condition is perturbed with 10\% ABC, and the flow achieves a magnetically dominated state with nearly maximal $H_M/E_T$. The three R17 runs, which all have 7\% ABC in both the velocity and magnetic field initial conditions, but are performed at different Reynolds numbers, evolve to a state on the boundary with both nonzero $H_M/E_T$ and $2H_C/E_T$ when given enough time. Note that although the three runs have the same initial conditions, they have take different paths through the parameter space with differing Reynolds number.

\section{The interplay between helical invariants} \label{s:HCM}
\subsection{Magnetic helicity relative growth}

According to the equations written in \S \ref{ss:stat}, the helical invariants play a central role in the evolution of MHD turbulence. Since the Taylor-Green flows have no helicity, we now examine a set of evolutions for several helical configurations
 that have been studied in the literature, namely the ABC (Beltrami) flows, the Orszag-Tang vortex and some perturbations of such flows
 (see \S \ref{ss:runs} and Table \ref{tab1} for definitions).
 
 The Orszag-Tang vortex, without magnetic helicity, becomes highly correlated, but with the inclusion of some magnetic helicity, it evolves toward states which, as $H_M/E_T$ increases, are more and more magnetically dominated. With very small additions of magnetic helicity (runs R23a and R23b), $H_M/E_T$ grows to modest values at which $2H_C/E_T$ can still obtain a value of one and the flow has equipartition between kinetic and magnetic energy. However, with even a slightly larger addition of $H_M$ (see runs R24 and R25 of Table \ref{tab1}), the growth of $H_M/E_T$ begins to dominate and the flow moves towards more magnetically dominated states.
 Similarly, unperturbed ABC flows, with strong $H_M$, remain uncorrelated if initially so; but when perturbing them by adding some correlation between the velocity and the magnetic field, as in run R20, they follow similar evolutions but stay away from the singularity that occurs at maximum $H_M/E_T$, $H_C\equiv 0$.

 \begin{figure*}  \begin{center}
\includegraphics[width=5.91cm   ]{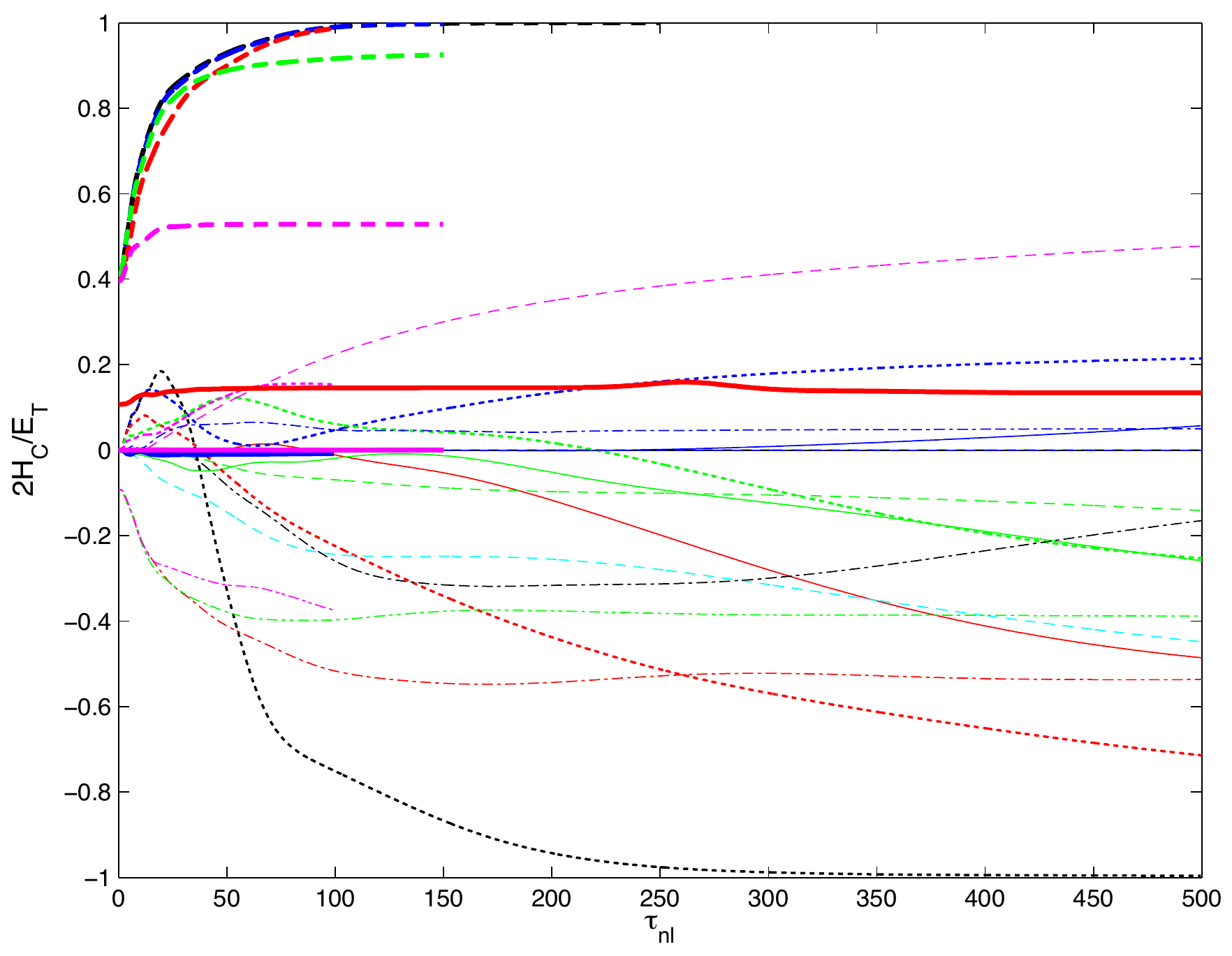}
\includegraphics[width=5.91cm ]{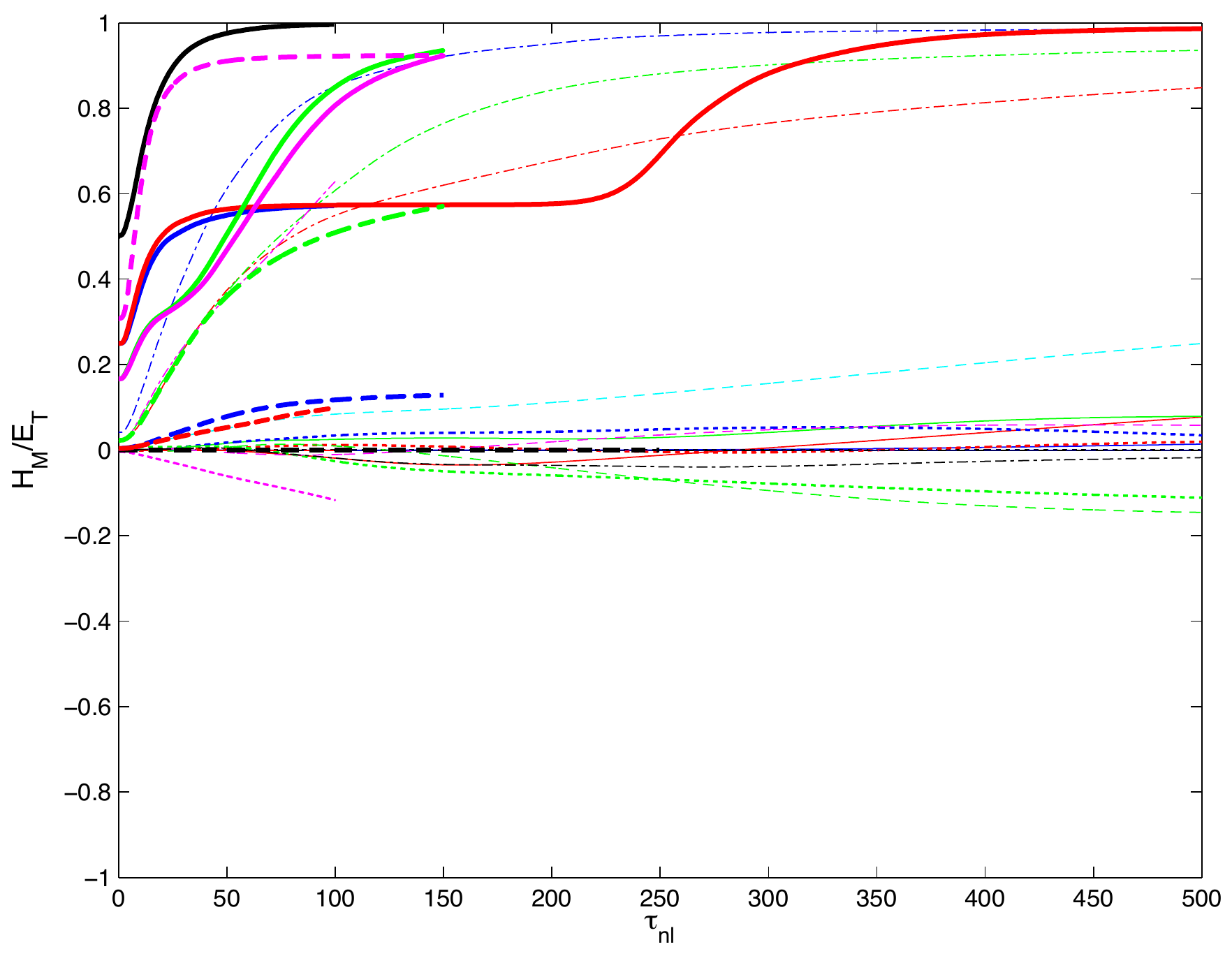}
\includegraphics[width=5.91cm ]{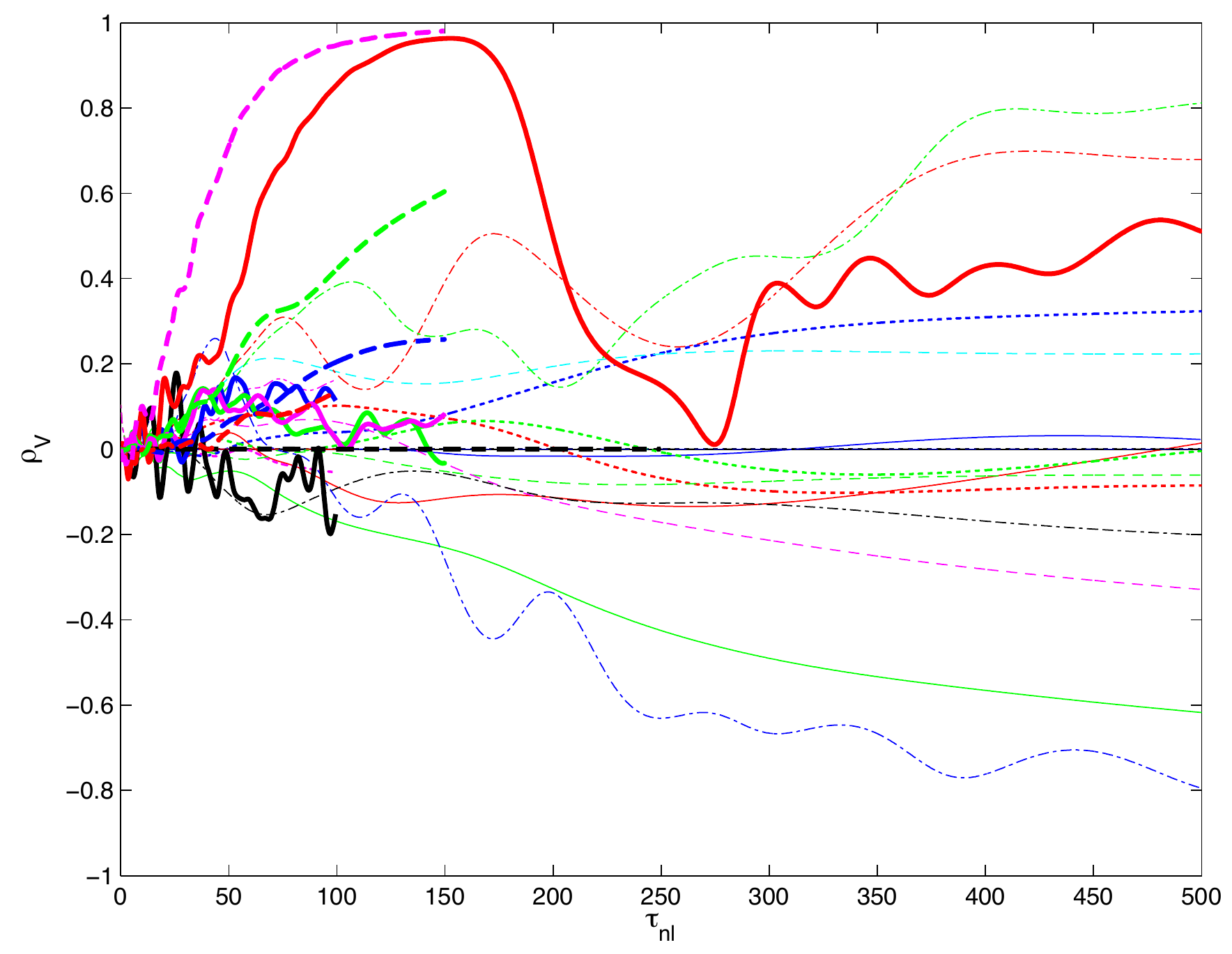}
 \caption{Temporal evolution of helicity in relative terms for all the runs of Table \ref{tab1}, with 
 $H_C/E_T$ (left),  $k_{min}H_M/E_T$ (middle) and $\rho_V$  (right, see definition in eq. (\ref{eq:relH})). The symbols and color scheme are the same as that in Fig. \ref{fig:ratio}. Note the clear distinction between the runs for the two invariant quantities (either staying at values close to zero or close to their extrema), and the more varied evolution for $\rho_V$ (right), although its evolution is also influenced by magnetic helicity (see eq. (\ref{eq:HJ})).
 }  \label{fig:rho}  \end{center}    \end{figure*}

 \begin{figure*} \begin{center}
\includegraphics[width=4.cm   ]{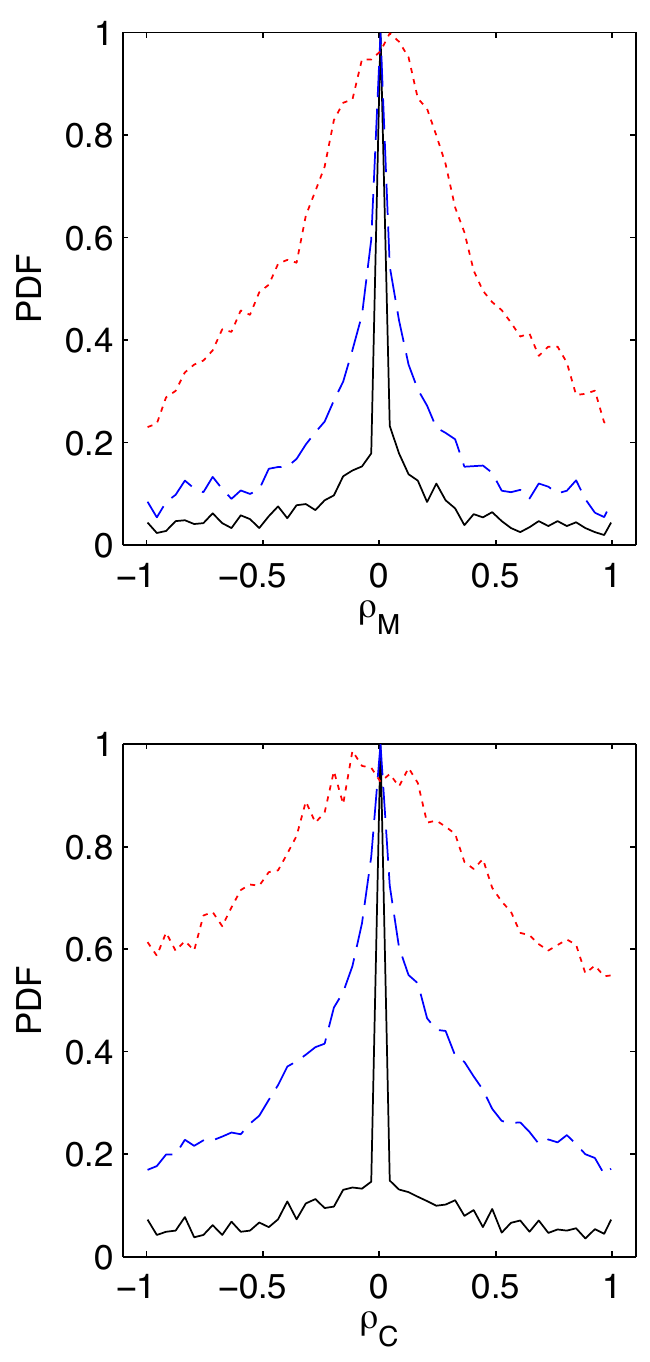}
\includegraphics[width=4.cm   ]{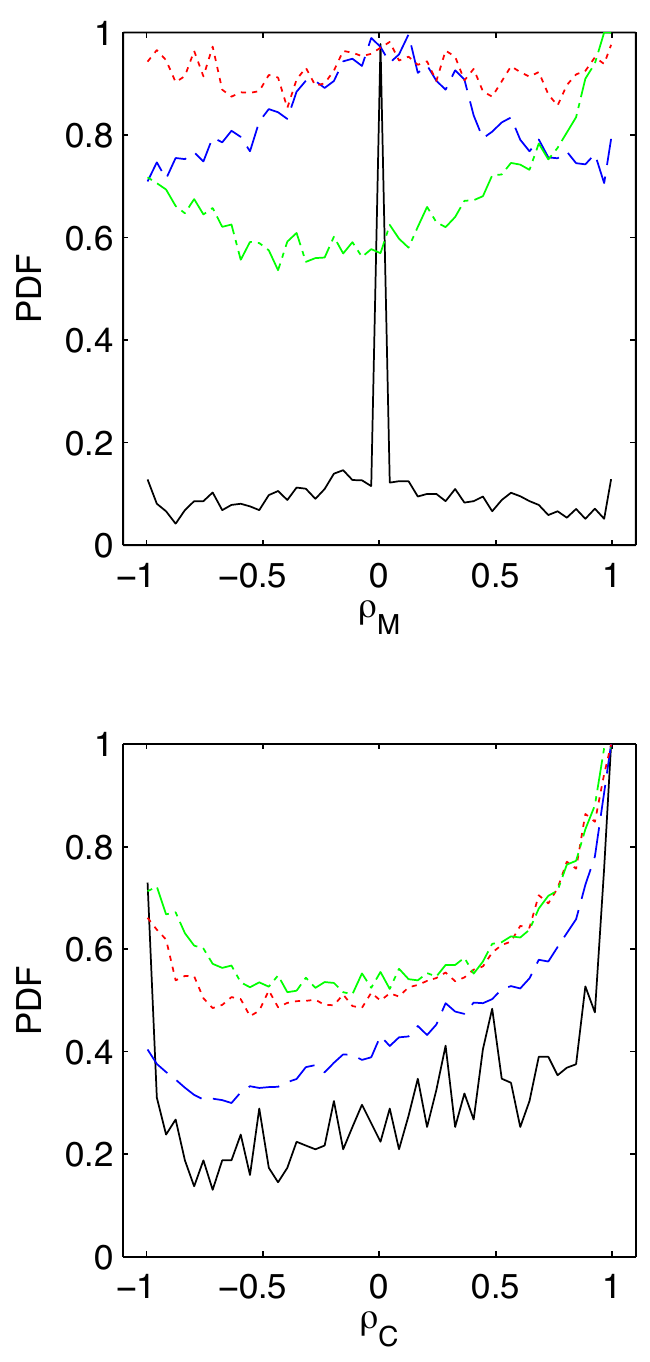}
\includegraphics[width=4.cm   ]{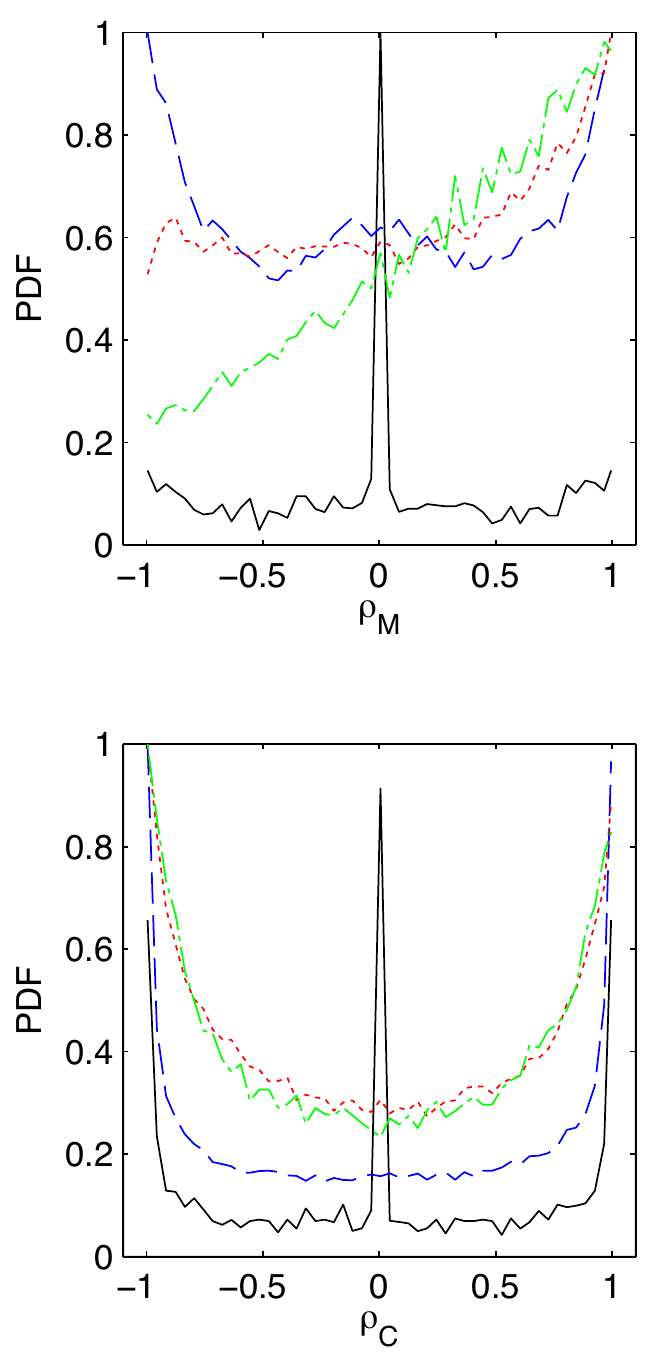}
\includegraphics[width=4.cm   ]{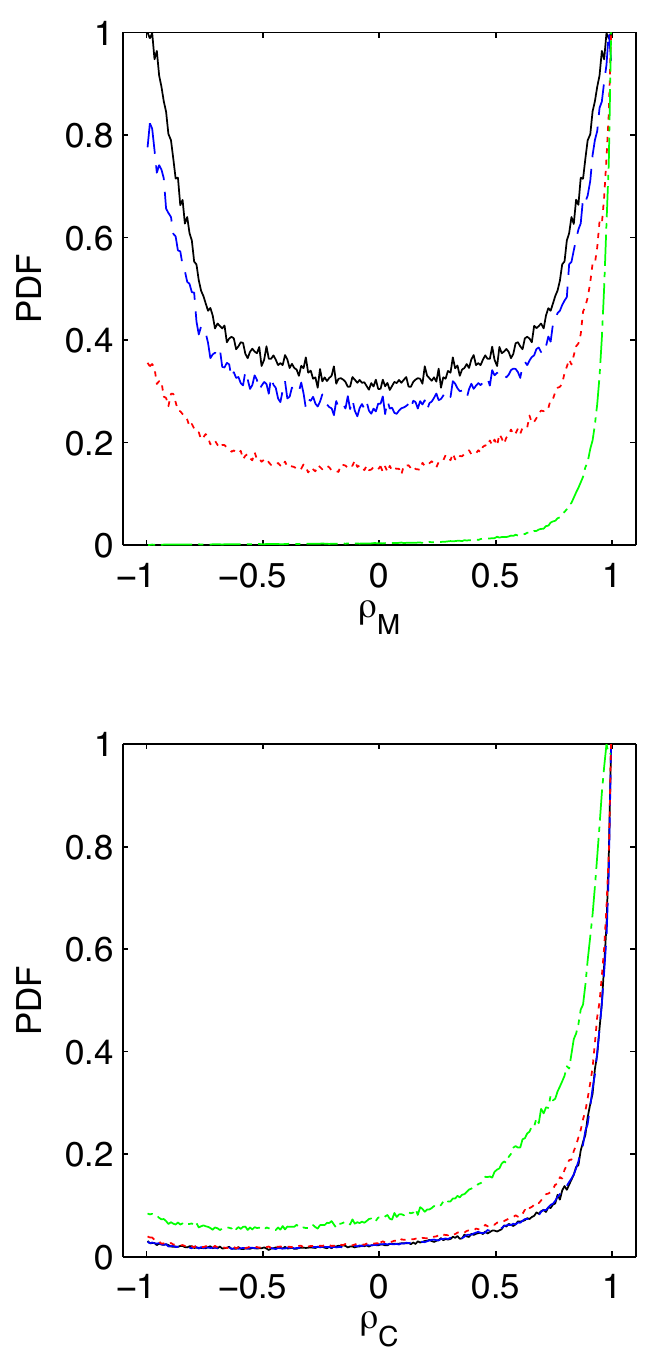}
 \caption{Probability distribution functions at $T=10\tau_{NL}$ of the cosine of the angle between the magnetic potential and the magnetic field (top row) and the velocity and magnetic field (bottom row), with all PDFs  normalized to their maximum value.
 {\it Left-most:} I flow runs R1a (black solid line), R2 (blue dashed line), and R3 (red dotted line);
  {\it Middle left:} C flow runs R5a (black solid line), R7 (blue dashed line), R8 (red dotted line), and R9a (green dashed-dotted line);
  {\it Middle right:} A flow runs R10a (black solid line), R12 (blue dashed line), R14 (red dotted line), and R16 (green dashed-dotted line);
 {\it Right-most:} OT flow runs R22 (black solid line), R23a (blue dashed line), R24 (red dotted line), and R25 (green dashed-dotted line); see Table \ref{tab1} for the nomenclature of the runs. 
  }  \label{fig:pdf} \end{center}  \end{figure*}


\subsection{Vector alignment in MHD turbulence}
Relative alignment of dynamical fields (see eq. (\ref{eq:relH}) for definitions) are shown in Fig. \ref{fig:rho} for many runs, using the same line (color) encoding as in the preceding figures. A lot of runs reach an  Alfv\'enic states ($\rho_C\sim \pm 1$), some more slowly, and a few stay at low values: it has been known for a long time that the correlation between the velocity and the magnetic field grows with time (see e.g. \cite{houches}).

Magnetic helicity seems more discriminating insofar as the long-time behavior of the runs: either $k_{min}H_M/E_T$ remains rather low, or else it approaches its maximal value.
The runs that approach near maximal values of $H_M/E_T$ are those in which a sufficient amount of magnetic helicity is present in the initial condition. In Fig. 2, as well as in Fig.  4 it, can be seen that all the runs which attain large values of $H_M/E_T$ have at least slightly larger initial values of magnetic helicity than those that remain near zero.
On the other hand, the normalized kinetic helicity shows a more varied set of behavior, with sometimes strong fluctuations between aligned and perpendicular fields, as for Run {R20} (thick solid, red line)
which evolves towards strong magnetic helicity (see the map in Fig. \ref{fig:map}). Note that Run R25 (thick dash, purple, line) is likely to evolve in a similar manner. This analysis suggests that one ought to look more in detail at the alignment properties of the various fields, by examining PdFs of the angle between various vectors. We show in Fig. \ref{fig:pdf} alignment probability distribution functions  
for several variables for several flows (see caption), after 10 turn-over times: the top row is for  magnetic potential and magnetic field, i.e. we are examining magnetic helicity, whereas the
 bottom row is for the velocity and magnetic field, i.e. we are concentrating then on the cross helicity.
  At $T=0$, all undisturbed flows have a strong central peak corresponding to orthogonality of vectors  (either ${\bf a}\perp {\bf b}$ or 
 ${\bf v}\perp {\bf b}$, and thus strong non-linearities, except for the OT case (right) for which  
 $\rho_M$ peaks symmetrically at values slightly greater and slightly less than zero  and $\rho_C$ indicates that there is a significant fraction of highly aligned velocity and magnetic field vectors.
 
These PDFs  
confirm the results illustrated in Fig. \ref{fig:map} in showing an  
evolution towards either alignment of the velocity and the magnetic  
field, or of the magnetic field and the potential, once the flows are  
perturbed, the more so the larger the perturbation.
The I flows (left-most column) are in fact the harder to perturb insofar as alignment does not really develop and one only observes a widening of the PdFs around zero, i.e. a distribution of angles that remain nevertheless close to $\pi/2$.
For the C flow family of runs (middle left column), a perturbation at the level of $10^{-x}$ simply widens the distribution of angles (blue dashed curve), but increasing this perturbation leads to a totally different behavior and a flat distribution for magnetic helicity, the perturbations being less significant for cross-helicity in the sense that the PdF is changed but the overall distribution (its shape) is similar in all cases..
The A flows (middle-right column), as the flow is more perturbed from its highly symmetric initial conditions, the fields become more aligned, with an almost equal distribution for ${\bf v}$ and ${\bf b}$ ($\rho_C\sim \pm 1$), whereas a clear alignment develops for ${\bf a}$ and ${\bf b}$ ($\rho_M\sim \pm 1$). Finally, the Orszag-Tang flow (right-most column) starts from a different configuration of vectors, and its evolution as it is more perturbed is 
not so dramatically different (except for the green dash-dotted line which has a 50\% OT-50\% ABC mixture).

On the other hand, the relative kinetic helicity (corresponding to alignment of velocity  
and vorticity) does not seem to follow a clear organization, unless  
magnetic helicity is strong and kinetic helicity does follow $H_M$ and grows in relative terms, under  
the influence presumably of Alfv\'en waves due to the large-scale  
magnetic field, as predicted in \cite{strong}. Growth of ${\bf v}-{\mathbf {\omega}}$ or ${\bf v}-{\bf b}$ alignment is a dynamical property of   
the Navier-Stokes or MHD equations, corresponding to the mutual  
interactions of shear and vorticity or shear and magnetic field  \cite{matthaeus_08}; in fact, such alignments properties have been  
found between all relevant fields, to different degrees  
\cite{servidio_08}. This alignment property is shared by the rather low Reynolds  
number computations performed here, and for long times as well.

\subsection{Is there a dynamically significant ratio in MHD turbulence?}
We finally examine the relative role of the velocity and the magnetic field, in terms of energy distribution and timescales,  for the runs performed at the highest Reynolds number, and thus with the largest extent of the inertial range (and numerical grid resolution). In order to do so, we look at the behavior  
in the inertial range around the peak of dissipation, when the  
turbulence is developed and the Reynolds number has not decreased  
substantially yet, of the ratio of magnetic to kinetic energy  
$r_E(k)=E_M(k)/E_V(k)$ and of the eddy turn-over time to the  
Alfv\'en time  $r_{\tau}(k)=\tau_{NL}(k)/\tau_A(k)$, the latter being built on the magnetic field in the gravest  mode. 

\begin{figure*} \begin{center}
\includegraphics[width=8.cm   ]{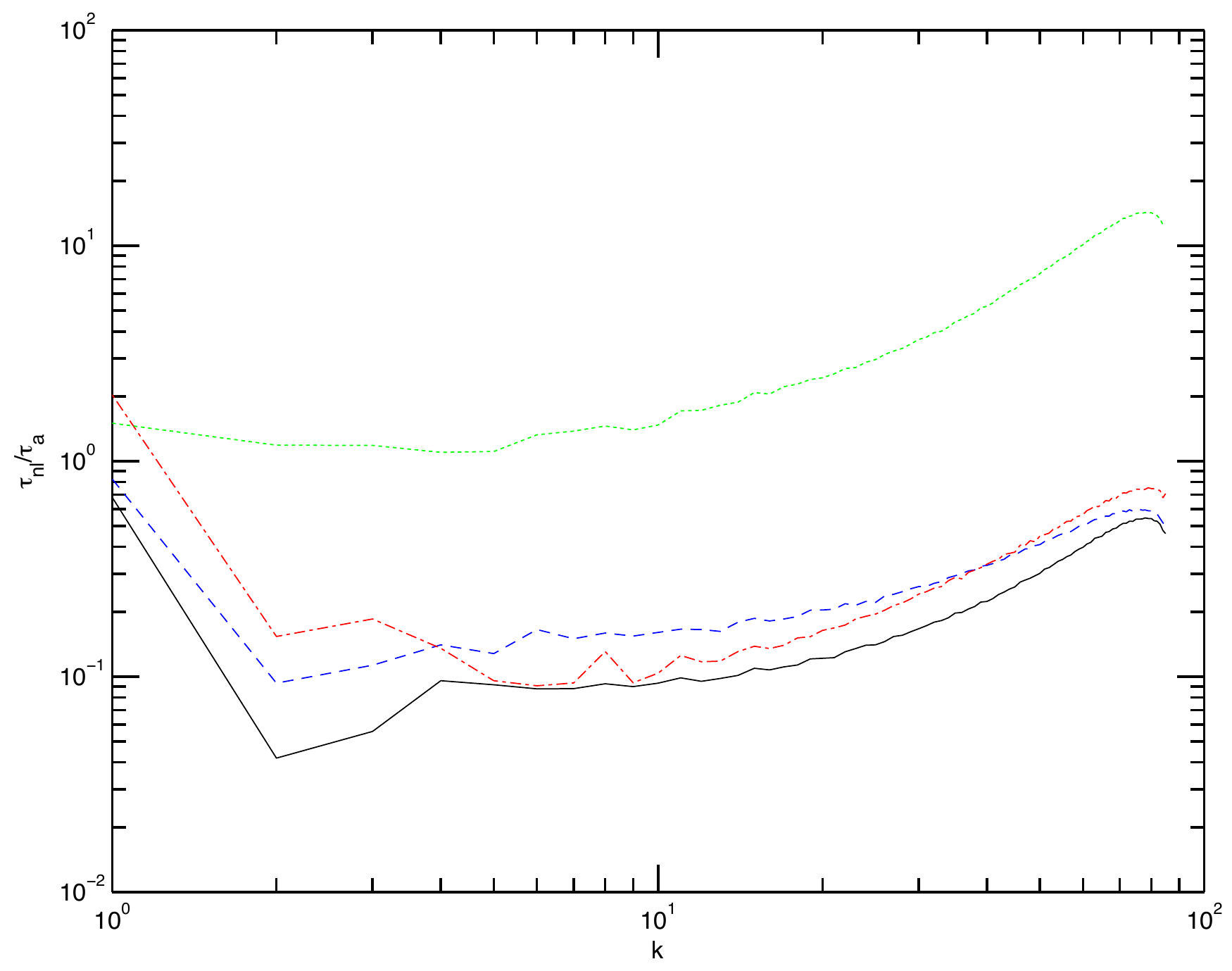}
\includegraphics[width=8.cm   ]{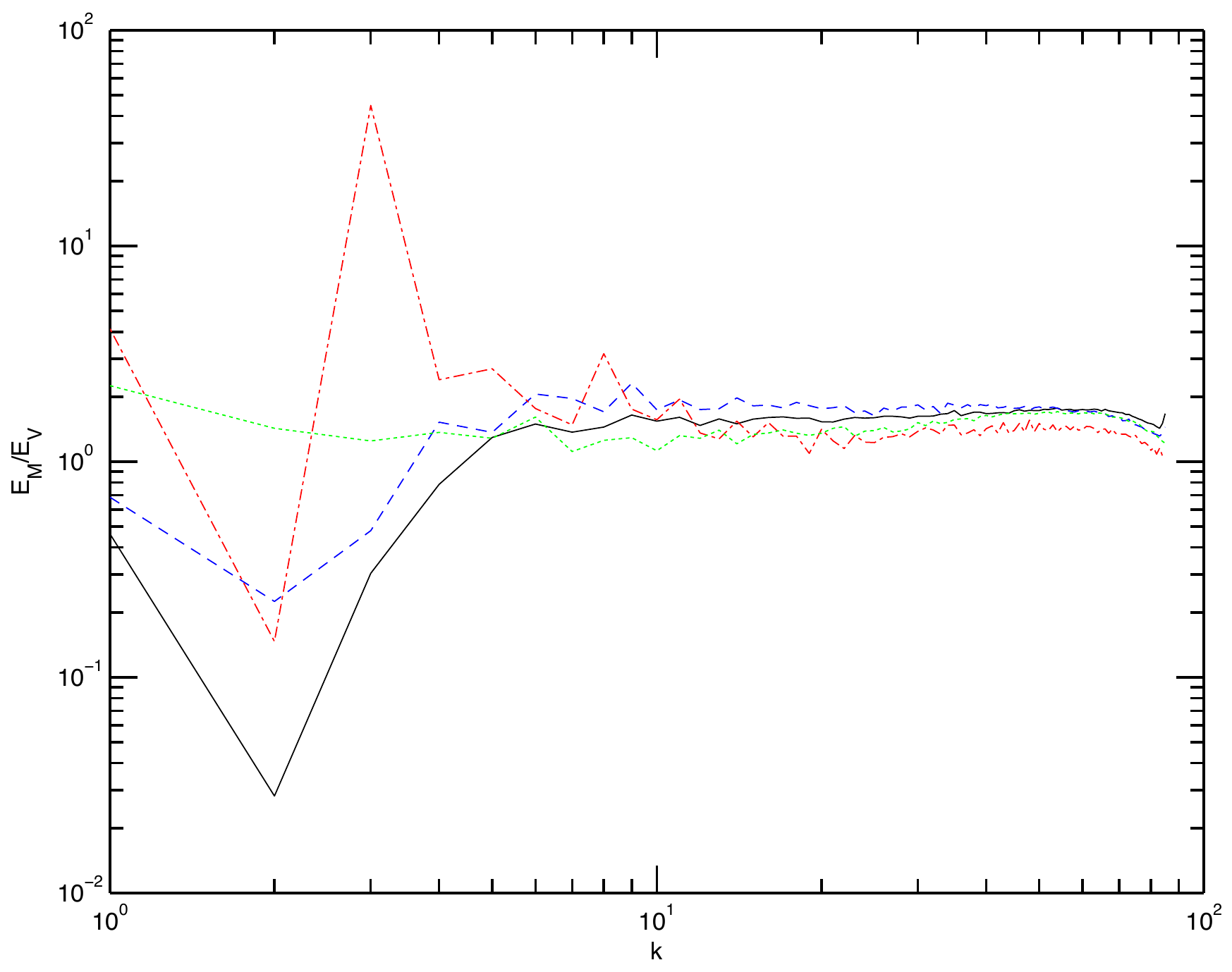}
  \caption{
  Nonlinear to Alfv\'en time  ratio {\it (left)} and magnetic to kinetic ratio {\it (right)}, as a function of wavenumber at peak of dissipation
  for the following computations performed on grids of $256^3$ points (see Table \ref{tab1} for the nomenclature):
    run R9b (C flow with strong perturbation, {\it black solid line}),
   run 17c (A flow with 7\% ABC, {\it blue dashed line}),
   run R21b (TG velocity and ABC magnetic field, {\it red dashed-dotted line}),
  and run R23b (99\% OT and 1\% ABC, {\it green dotted line}).
   Note the constancy of quasi-equipartition of energy throughout the inertial  
range, to be contrasted with the increase in the ratio of  
characteristic time scales in that same range.
   }  \label{fig:EMsEV} \end{center}  \end{figure*}

The ratio of timescales behaves as expected when evaluating the turn-over time on the energy spectrum variation with wavenumber.
This can be seen in Fig. \ref{fig:EMsEV}: for all flows, $r_{\tau}(k)$ increases with decreasing scale because of  
the way these two characteristic times change with scale, i.e. $1/k$  
for $\tau_A$ and $[k^3E_V(k)]^{-1/2}$ for the eddy turnover time. On  
the other hand, and again for all flows, the energy ratio is constant and of order unity  
(but systematically slightly above in fact, as also regularly observed often in the Solar Wind), except in the largest scale  
in which it is dominated by initial conditions and the reinforcement  
of magnetic energy in the largest scale in all the runs dominated by  
an evolution towards the top of the map displayed in Fig.  
\ref{fig:map}, and corresponding to cases with strong magnetic helicity. We also  
note that at later times of the order of three times the peak of  
dissipation, this result still holds but with, in all cases displayed  
here, an increase in $E_M/E_V$ at the largest excited {scales}, by  a  factor of 2 to 40 (not shown).
All other runs of this study behaved similarly, as long as the Reynolds number is sufficiently high for turbulent mixing to take place.

\section{Conclusions}\label{s:conclu}
We have shown in this paper that the Taylor-Green configurations studied in \cite{lee} for their energetic properties, depart from their strong symmetries given a strong-enough perturbation, as can be encountered in high Reynolds number flows. They then evolve toward different characteristic behaviors in the $H_M, \ H_C$ plane (map of Fig. \ref{fig:map}); these can depend on the perturbation (whether it has cross-correlation or magnetic helicity, leading the flows to different end-states in particular). It is not clear however if, when studying these flows at substantially higher Reynolds numbers, as was done in \cite{lee} but, contrary to \cite{lee},  not imposing the symmetries at all time, one will still have three different scaling laws for the total energy spectra for these three configurations.
We also confirm that statistical mechanics, with an energy minimization principle, is an excellent predictor for the behavior of turbulent flows, as argued in \cite{stribling_90}, and studied at moderate resolutions in \cite{stribling_91}, but the question still remains as to whether this leads to different scaling laws in MHD turbulence at high Reynolds numbers.

Quasi-equipartition between the kinetic and magnetic energy is expected, on the basis of mixing of complex systems with a large number of degrees of freedom, although as shown in \cite{frisch_75}, magnetic helicity, alone or in the presence of cross-helicity, may well prevent this from happening. What we have shown in this paper is that, in some cases with strong phase relationships such that the nonlinear terms are weakened considerably through alignment of the relevant fields  (vorticity, velocity, magnetic field, magnetic current), other solutions are reachable with quite different properties. This is a bit akin to a potential flow in hydrodynamics: when the vorticity is identically zero, this is an exact solution of the Navier-Stokes equations although an unstable one, and vorticity, like a seed magnetic field, grows over time.
The reason why these solutions do not destabilize in a Lyapounov time, which can be close to an eddy turn-over time, is probably due to the fact that symmetries are very strong properties of flows that are preserved by the dynamical evolution \cite{falko_sym, bardos}; would stronger turbulent flows at higher Reynolds number lead to more complex behavior than what is observed here? Or, since  MHD turbulence is known to be a dissipative system including in the limit of infinite Reynolds number \cite{mininni_finite_eps}, 
 is the emergence of large-scale helical coherent structures enough to insure a simple evolution following an energy minimization principle?


\begin{acknowledgments}  
Help in start-up runs from Duane Rosenberg is gratefully acknowledged. The National Center for Atmospheric Research where the computations were performed is sponsored by the National Science Foundation. This material is based upon work supported by the National Science Foundation Graduate Research Fellowship under Grant No. DGE 1144083 for Joshua Stawarz. Finally,  Marc Brachet acknowledges a CISL fund allocation. 
\end{acknowledgments}

\end{document}